\RequirePackage{fix-cm}
\documentclass[smallextended]{svjour3}       
\smartqed  
\usepackage{graphicx}
\usepackage{amssymb}
\usepackage{amsmath}
\usepackage{subfigure}
\newcommand{\CSSM}{Special Research Centre for the Subatomic Structure
  of Matter (CSSM),\\Department of Physics, University of
  Adelaide, Adelaide, South Australia 5005, Australia}

\newcommand{\CoEPP}{ARC Centre of Excellence for Particle Physics at
  the Terascale (CoEPP),\\Department of Physics, University
  of Adelaide, Adelaide, South Australia 5005, Australia}

\newcommand{\Lanzhou}{School of Physical Science and Technology,\\ Lanzhou University, Lanzhou 730000, China }

\newcommand{\KEK}{KEK Theory Center, Institute of Particle and Nuclear Studies (IPNS), High Energy Accelerator
Research Organization (KEK), Tsukuba, Ibaraki 305-0801, Japan }

\newcommand{\JPAC}{J-PARC Branch, KEK Theory Center, IPNS, KEK, Tokai, Ibaraki 319-1106, Japan }

\newcommand{\Argonne}{Physics Division, Argonne National Laboratory, Argonne, Illinois 60439, USA}

\newcommand{\Bonn}{HISKP (Theory ) and BCTP, University of Bonn, Germany}

\begin{document}

\title{Nucleon Excited States from Lattice QCD and Hamiltonian Effective Field Theory
}

\titlerunning{Nucleon Excited States from LQCD and HEFT}        

\author{ Jia-jun Wu         \and
                     Jonathan M. M. Hall \and
                      H. Kamano \and 
                      Waseem Kamleh \and 
                      T.-S. H. Lee \and 
                      Derek B. Leinweber \and 
                      Zhan-Wei Liu \and  
                      Finn M. Stokes \and 
                      Anthony W. Thomas
}


\institute{ Jia-jun Wu \at
             \CSSM \\ \Bonn \\
              \email{jiajun.wu@adelaide.edu.au}           
           \and
            Jonathan M. M. Hall,\, Waseem Kamleh,\,  Derek B. Leinweber,\,  Finn M. Stokes,\,   Anthony W. Thomas \at
             \CSSM \\ \CoEPP
           \and
           Zhan-Wei Liu \at
             \Lanzhou
             \and
              H. Kamano  \at
              \KEK \\ \JPAC
             \and
             T.-S. H. Lee  \at
             \Argonne              
}

\date{Received: date / Accepted: date}

\maketitle

\begin{abstract}
An approach for relating the nucleon excited states extracted from lattice QCD and the nucleon resonances of experimental data has been developed using the Hamiltonian effective field theory (HEFT) method.  By formulating HEFT in the finite volume of the lattice, the eigenstates of the Hamiltonian model can be related to the energy eigenstates observed in Lattice simulations.  By taking the infinite-volume limit of HEFT, information from the lattice is linked to experiment. The approach opens a new window for the study of experimentally-observed resonances from the first principles of  lattice QCD calculations.  With the Hamiltonian approach, one not only describes the spectra of lattice-QCD eigenstates through the eigenvalues of the finite-volume Hamiltonian matrix, but one also learns the composition of the lattice-QCD eigenstates via the eigenvectors of the Hamiltonian matrix.  One learns the composition of the states in terms of the meson-baryon basis states considered in formulating the effective field theory. One also learns the composition of the resonances observed in Nature.   In this presentation, I will focus on recent breakthroughs in our understanding of the structure of the $N^*(1535)$, $N^*(1440)$ and $\Lambda^*(1405)$ resonances using this method.
\end{abstract}

\section{Introduction}
\label{sec:intro}

The nature of the low lying resonances is one of the most challenging problems in modern hadron physics.
It is still hard to understand the spectra of light baryons and mesons within Quantum Chromodynamics (QCD) because of the non-perturbative property.
Lattice QCD realizes the non-perturbative calculation based on the Mentor Carlo simulation technique in the finite-volume.
In recent years, it makes a lot of important progresses~\cite{Mohler:2017ibi}.
It is therefore necessary to connect the lattice QCD results and experimental data to investigate the resonances.
There are various approaches to connect the experimental data and the energy levels in the finite-volume \cite{Luscher:1986pf,Luscher:1990ux,Rummukainen:1995vs,Kim:2005gf,He:2005ey,Hansen:2012tf,Briceno:2013lba,Gockeler:2012yj,Briceno:2014oea,Doring:2011vk,Doring:2012eu,MartinezTorres:2011pr,Doring:2011nd,Doring:2013glu,Geng:2015yta}.

In this report,  the Hamiltonian effective field theory (HEFT) is reviewed based on recent references, \cite{Hall:2013qba,Wu:2014vma,Liu:2015ktc,Hall:2014uca,Liu:2016uzk,Liu:2016wxq,Wu:2017qve,Wu:2016ixr}.
HEFT is an extension of chiral perturbation theory~\cite{Hall:2013qba,Young:2002ib}.
In the infinite volume, through solving a three-dimensional reduction of the coupled-channel Bethe-Salpeter equation, the Hamiltonian generates the scattering amplitudes, T-matrix, which is straightforward to the experimental observables.
On the other hand, through solving a eigen-equation of the finite-volume Hamiltonian matrix, the eigen-values corresponding to the energy levels of lattice QCD can be obtained.
From one Hamiltonian model, we build the relationship between the experimental data and finite-volume energy levels.
This relationship is consistent with the model independent formalism developed by L\"uscher~\cite{Luscher:1986pf,Luscher:1990ux} as proved in Ref.\cite{Wu:2014vma}.
Although HEFT relies on the Hamiltonian model, the model dependence parts in this relationship  are absorbed into the exponential suppressed terms of the lattice size. 
Furthermore, HEFT also provides the eigenvectors of the eigenstates in the finite-volume.
The eigenvectors directly revel the composition of these eigenstates which carry information on the internal structure of the resonances.
Therefore, HEFT provides an unique way to study the inside structure of resonances with experimental and lattice QCD data.

In principle, HEFT requires to measure all eigenstates in the finite-volume, however, current lattice techniques can not arrive this requirement.
For the baryon pure three-quark local operators are mostly used, while rarely some calculations use Meson-Baryon non-local operators~\cite{Lang:2016hnn}.
In this report, we focus on the results from three-quark local operators.
The reasonable assumption is that the state extracted in the finite-volume by three-quark local operators should take a large component of three-quark core.
HEFT can be used to examine the three-quark core components of the $N^*(1535)$, $\Lambda^*(1405)$ and $N^*(1440)$ states \cite{Liu:2015ktc,Hall:2014uca,Liu:2016uzk,Liu:2016wxq,Wu:2017qve}.
In this report, we review recent results and build a new picture of low lying resonances.

The framework of HEFT is described in Sec. \ref{sec:from}.
The relationship between the eigenstates in the finite-volume and the resonances are discussed. 
In Sec.\ref{sec:results}, the numerical results and discussions of $N^*(1535)$, $\Lambda^*(1405)$ and $N^*(1440)$ are presented.
Based on the studies of these resonances, a new picture of the low lying hadron spectrum is proposed.
A brief summary concludes in Sec. \ref{sec:con}.

\section{Framework of HEFT}
\label{sec:from}

\subsection{The Hamiltonian}

In the rest frame, the Hamiltonian has the following form
\begin{eqnarray}
H = H_0 + H_I \, ,
\label{eq:h}
\end{eqnarray}
where the non-interacting Hamiltonian is
\begin{eqnarray}
H_0 &=&\sum_{B_0} |B_0\rangle \, m^{0}_{B} \, \langle B_0|+ \sum_{\alpha}\int d^3\vec{k}
|\alpha(\vec{k})\rangle \, \left [\, \omega_{\alpha_M}(\vec{k})
+ \omega_{\alpha_N}(\vec{k})\, \right ] \langle\alpha(\vec{k})| \, .
\label{eq:h0}
\end{eqnarray}
Here $B_0$ denotes a bare baryon with mass $ m^{0}_{B}$, which may be thought of
as a quark model state, $\alpha$ designates the
channel and $\alpha_M$ ($\alpha_N$) indicates the meson (baryon) state which
constitutes channel $\alpha$. The energy $\omega_{\alpha_i}(\vec{k})
= \sqrt{m^2_{\alpha_i}+\vec{k}^2}$.

The energy independent interaction Hamiltonian includes two parts, $H_I = g + v$,
where $g$ describes the vertex interaction between the bare particle
and the two-particle channels $\alpha$
\begin{eqnarray}
g &=& \sum_{\alpha,\, B_0} \int d^3\vec{k} \left \{  \,
|\alpha(\vec{k})\rangle \, G^\dagger_{\alpha, B_0}(k)\, \langle B_0|+h.c.
\right \}\, ,
\label{eq:int-g1}
\end{eqnarray}
while the direct two-to-two particle interaction is defined by
\begin{eqnarray}
v = \sum_{\alpha,\beta} \int d^3\vec{k}\; d^3\vec{k}'\,
|\alpha(\vec{k})\rangle\,  V^{S}_{\alpha,\beta}(k,k')\, \langle
\beta(\vec{k}')| \, .
\label{eq:int-v}
\end{eqnarray}

Here we consider three different resonances, $N^*(1535)$, $N^*(1440)$ and $\Lambda^*(1405)$, so different interaction forms and channels for them are involved.
For the $N^*(1535)$, it includes $\pi N$ and $\eta N$ channels, and $N^*(1440)$ includes  $\pi N$, $\pi \Delta $ and $\sigma N$ channels, while it has $\pi \Sigma$, $\bar{K} N$, $\eta \Lambda$ , $K \Xi$ and $\pi \Lambda$ channels for $\Lambda^*(1405)$.

For the vertex interaction between the bare baryon and the two-particle
channels we choose:
\begin{eqnarray}
G_{\alpha, B_0}^2(k)&=&\frac{3\,g_{B_0\alpha}^2}{4\,\pi^2\, f^2}\,\omega_{\alpha_1}(k) u^2(k) \, ,
\end{eqnarray}
for $N^*(1535)$ and $\Lambda^*(1405)$ case,
where  the pion decay constant $f=92.4$ MeV and the form factor is chosen as  $u(k)=\left(\, 1+ k^2/ \Lambda^2\,\right )^{-2}$, and $\Lambda = 0.8$ and $1.0$ GeV for $N^*(1535)$ and $\Lambda^*(1405)$, respectively.
For the  $N^*(1440)$ case, we choose: 
\begin{eqnarray}
G_{\alpha, B_0}^2(k)&=&\frac{g_{B_0\alpha}^2}{4\,\pi^2}\left (\frac{k}{f}\right )^{2l_\alpha} \frac{ u_{\alpha}^2(k)}{\omega_{\alpha_1}(k)} \, ,
\end{eqnarray}
where $l_\alpha$ is the orbital angular momentum in
channel $\alpha$. Here, since we are concerned with the Roper resonance, with isospin, angular
momentum and parity, $\mathbf{I(J^P)=\frac12(\frac12^+)}$, $l$ is 1 for $\pi N$ and $\pi \Delta$,
while it is 0 for $\sigma N$.  The regulating form factor, $u_\alpha(k)$, takes the exponential
form
$u_\alpha(k)=\exp\left ( -{k^2}/{\Lambda_\alpha^2} \right ) ,$
where $\Lambda_\alpha$ is the regularization scale.

For the direct two-to-two particle interaction, we introduce the separable
potentials for the $N^*(1535)$ and $N^*(1440)$.
For the $N^*(1535)$, only $\pi N \to \pi N$ interaction is applying, 
\begin{eqnarray}
V^S_{\pi N, \pi N}\,(\,k,\,k'\,)&=&\frac{3g^{N^*(1535)}_{\pi N, \pi N}\, \tilde{u}_{\alpha}(k)\, \tilde{u}_{\pi N}(k')}{4\,\pi^2\, f^2} \label{eqVspiN1535}
      \end{eqnarray}
where $ \tilde{u}_{\pi N}(k) = u(k)(m_\pi + \omega_{\pi}(k))/ \omega_{\pi}(k)$ and $u(k)$ is the dipole form used in the $G_{\alpha, B_0}$ coupling.
For the $N^*(1440)$,
\begin{eqnarray}
V^S_{\alpha, \beta}\,(\,k,\,k'\,)&=&g^{N^*(1440)}_{\alpha, \beta}\,\frac{\bar G_{\alpha}(k)}{\sqrt{\omega_{\alpha_1}(k)}}\,
      \frac{\bar G_{\beta}(k')}{\sqrt{\omega_{\beta_1}(k')}}\, ,\label{eqVspiN1440}
      \end{eqnarray}
where $\bar G_{\alpha}(k)=G_{\alpha, B_0}(k)/g_{B_0 \alpha}$.
For $\Lambda^*(1405)$, we use the Weinberg-Tomozawa term \cite{Veit:1984sf} as follows:
\begin{eqnarray}
V^S_{\alpha, \beta}\,(\,k,\,k'\,)&=&g^{\Lambda^*(1405)}_{\alpha, \beta}\,
\frac{\left[\,\omega_{\alpha_M}(k)+\omega_{\beta_M}(k')\,\right]u(k)\,u(k')}
{16\,\pi^2\,f^2\sqrt{\omega_{\alpha_M}(k)\omega_{\beta_M}(k')}}\, ,\label{eqVspiL1405}
\end{eqnarray}
where the form factor is same as that in the vertex between bare particle and two-particle channels.

\subsection{Infinite-Volume scattering amplitude}

The $T$-matrices for two particle scattering are obtained by solving
a three-dimensional reduction of the coupled-channel Bethe-Salpeter
equations for each partial wave
\begin{eqnarray}
t_{\alpha, \beta}(k,k';E)&=&V_{\alpha, \beta}(k,k';E)+\sum_\gamma \int
q^2 dq \times
\,
\frac{V_{\alpha, \gamma}(k,q;E)\,  t_{\gamma, \beta}(q,k';E)}{E-\omega_{\gamma_1}(q)-\omega_{\gamma_2}(q)+i \epsilon}\, 
\, . \nonumber
\end{eqnarray}
The coupled-channel potential is readily calculated from the interaction Hamiltonian
\begin{eqnarray}
V_{\alpha, \beta}(k,k') &=& \sum_{B_0}\,
\frac{G^\dag_{\alpha, B_0}(k)\, G_{\beta, B_0}(k')}{E-m_B^0} +V^S_{\alpha,\beta}(k,k') \, ,
\label{eq:lseq-2}
\end{eqnarray}
with the normalization $\langle \alpha(\vec{k})\, |\, \beta(\vec{k}^{\,\,'})\rangle =
\delta_{\alpha,\beta}\, \delta (\vec{k}-\vec{k}^{\,\,'})$.

Finally, the S-matrix is related to the
$T$-matrix by
\begin{eqnarray}
S_{\alpha, \beta}(E)&=&1-2i\, \sqrt{\rho_{\alpha}(E)\rho_{\beta}(E)}
\, t_{\alpha,\beta}(k_{\alpha\,\rm cm}, k_{\beta\,\rm cm}; E) \, ,\\
\rho_{\alpha}(E)&=& \pi\frac{\omega_{\alpha_M\,\rm cm}\, \omega_{\alpha_N\,\rm cm}}{E}\, k_{\alpha\,\rm cm} \, ,
\end{eqnarray}

where $k_{\alpha\,\rm cm}$ satisfies the on-shell condition $\omega_{\alpha_M\, \rm cm}+\omega_{\alpha_N\,\rm cm}=E$, and $\omega_{\alpha_{i}\,\rm cm} =\omega_{\alpha_{i}}(k_{\rm cm})$.

The pole position of any bound state or resonance is obtained by searching for the
poles of the $T$-matrix in the complex plane.

The cross section $\sigma_{\alpha\,,\beta}$ for the process $\alpha \to \beta$ is
\begin{eqnarray}
\sigma_{\alpha\,,\beta}&=&\frac{4\pi^3 k_{\alpha\,\rm cm}
\omega_{\alpha_M\,\rm cm}\, \omega_{\alpha_N\,\rm cm}\, 
\omega_{\beta_M\,\rm cm}\, \omega_{\beta_N\,\rm cm}}
{E^2\, k_{\beta\,\rm cm}}
|t_{\alpha,\beta}(k_{\alpha\,\rm cm}, k_{\beta\,\rm cm}; E)|^2 \,.
\end{eqnarray}

\subsection{The Finite-Volume Matrix Hamiltonian}

The formalism of the finite-volume matrix Hamiltonian is presented by following Refs.\cite{Hall:2013qba,Wu:2014vma}.
We can discretise the Hamiltonian in a box with length $L$. 
A particle can only carry momenta $k_n = \sqrt{n}\,2\,\pi /L$ in the box, where $n = 0$, $1$, $\cdots$.
The Hamiltonian in the finite-volume is the momentum discretization of the Hamiltonian $H$ at infinite volume.
Therefore, Eq.\,(\ref{eq:h}) can be extended in the finite-volume as:
\begin{eqnarray}
H^V &=& H^V_0 + H^V_I =H^V_0 +g^V+ v^V \, ,\label{eq:hL}
\\
H^V_0 &=&\sum_{B_0} |B_0\rangle \, m^{0}_{B} \, \langle B_0|
\nonumber\\&& 
\hspace{0.1cm}
+ \sum_{n}\sum_{\alpha}
|\alpha(k_n)\rangle_V \, \left [\, \omega_{\alpha_M}(k_n)
+ \omega_{\alpha_N}(k_n)\, \right ]\, _V\langle\alpha(k_n)| \, ,
\\
g^V &=&  \sum_{n} \sum_{\alpha,\, B_0} 
\sqrt{\frac{C_3(n)}{4\pi}}\left(\frac{2\pi}{L}\right)^{3/2} 
\left \{  \,
|\alpha(k_n)\,\rangle_V \, G^\dagger_{\alpha, B_0}(k_n)\, \langle B_0|+h.c.
\right \}\,, \\
 v^V &=&  \sum_{n, n'} \sum_{\alpha,\beta} 
\sqrt{\frac{C_3(n)\,C_3(n')}{16\,\pi^2}}\left(\frac{2\pi}{L}\,\right)^{3}|\alpha(k_n)\,\rangle_V\,  V^{S}_{\alpha,\beta}(k_n,k_{n'})\, _V\langle
\beta(k_{n'})| \, .
\end{eqnarray}
Here, $C_3(n)$ represents the number of ways of summing the
squares of three integers to equal n.
The normalization of $|\alpha(k_n)\,\rangle_V $ becomes $_V\langle \alpha(k_n)\,|\,\beta(k_{n'})\,\rangle_V = \delta_{nn'}\delta_{\alpha\beta}$.
Then it is straightforward to get the matrix of Hamiltonian in the finite-volume.
One can obtain the eigenstate energy levels on the lattice and analyse the corresponding eigenvector wave functions describing the constituents of the eigenstates with the finite-volume Hamiltonian matrix.

In addition to the results at physical pion mass, we can also extend the formalism to unphysical pion masses.
Using $m^2_\pi$ as a measure of the light quark masses, we consider the variation of the bare ($B_0$) mass and meson ($M$) mass as:
\begin{eqnarray}
m_B^0(m_\pi^2)&=&m_B^0|_{\rm phy}+\alpha_{B_0}\, (m_\pi^2-m_\pi^2|_{\rm phy})\, , \\
m_M^2(m_\pi^2)&=&m_M^2|_{\rm phy}+\alpha_M\, (m_\pi^2-m_\pi^2|_{\rm phy}),
\end{eqnarray}
where the slope parameters $\alpha_{B_0}$ for bare states are constrained by lattice
QCD data from the CSSM. 
The stable particles such as $N$, $\Delta$, $\Sigma$ and $K$ use linear interpolations between the corresponding lattice QCD results.
For the $\sigma$, it uses $\alpha_{\sigma} = \frac{4}{3}m_{\sigma}|_{\rm phy} \alpha_N$ = 0.67\cite{Cloet:2002eg}, where  $ \alpha_N = 1.0$ GeV$^{-1}$.
The slope parameter of the $\eta$ meson is used $\frac{1}{3}$.

\subsection{The relationship between the eigenstate and the resonance }

\begin{figure}[tbp] \vspace{-0.cm}
\begin{center}
\includegraphics[width=1.0\columnwidth]{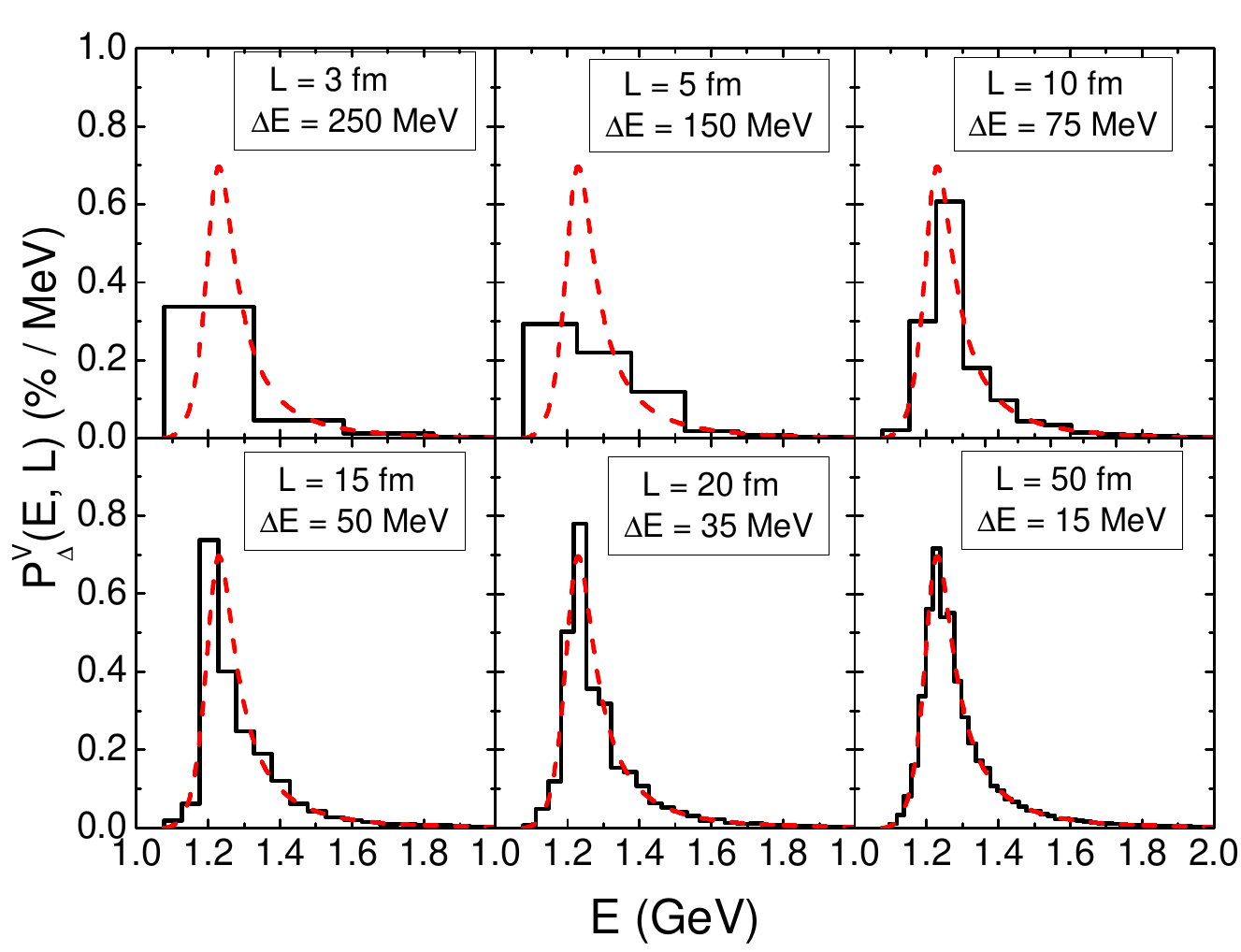}
\caption{The probability $P^V(E, L)$ (black solid line) and $P(E)$ (red
  dashed curve) of the Sato-Lee model model at the $\Delta(1232)$ energy region with $L=3, 5, 10, 15, 20,$ and 50 fm.}
\label{fg:delta}
\end{center}
\end{figure}

The main difference between finite and infinite volume in the momentum space is the discretization of the momentum in the box.
It leads to the discrete eigenvalue of the Hamiltonian matrix.
Therefore, the eigenstates in the box are the discrete states of the continuum final scatting states in the infinite-volume. 
Each eigenstate can not be recognized as a resonance directly, while the resonance is recognized as a pole position on the complex plan of the scattering amplitude. 
As we known, the resonance will make enhancement of the cross-section, i.e., the bump on the figures of  cross-section vs scattering energy.
In other words, when we find one peak of cross-section vs total energy, sometimes it may exists a resonance around this energy.
Around one peak, there are infinite final scattering states, the behavior of resonance at the real axis is taken by these states.
In other words, the pole position on the complex plan will be shown as a peak on the real axis.
Such resonance peak can be also observed by the overlap between bare state and final scattering state, as shown the red color line in Fig.\ref{fg:delta}.
Correspondingly, the overlap between bare state and eigenstate of finite volume also can be used to find such peak for resonance.
Based on Ref.~\cite{Wu:2016ixr}, we define two variables $P^V$ and $P$ for finite and infinite volume, respectively, 
\begin{eqnarray}
P^V(E^{ave}_k,L\,)& =&\frac{1}{Z^V}\frac{1}{\Delta E} 
\sum_{E^{ave}_k-\frac{\Delta E}{2} \leq E_i\leq E^{ave}_k+\frac{\Delta E}{2}} 
|\,\langle B_0 |\, \Psi^V_{E_i}\,\rangle\,|^2 \,,
\label{eq:PEL}
\\
P(E\,)&=&\frac{1}{Z}\sum_{\alpha} \pi\,k_{\alpha\,\rm cm}\,\omega_{\alpha_M\,\rm cm}\,\omega_{\alpha_B\,\rm cm}
|\,\langle B_0|\,\Psi^{(+)}_{E,\alpha}\,\rangle\, |^2\,,
\label{eq:PE}
\end{eqnarray}
where $Z^V$ and $Z$ are the normalization factors, and $|\Psi^{(+)}_{E,\alpha}\,\rangle$ and $ | \Psi^V_{E_i}\,\rangle$ are the finial scattering states of channel $\alpha$ in the infinite-volume and $i-th$ finite-volume eigenstates, respectively.

As an example, we use the Sato-Lee model~\cite{Sato:1996gk,Sato:2000jf,JuliaDiaz:2006xt} to calculate the Hamiltonian including the bare $\Delta$ state and $\pi N$ channel.
The eigenstates in the finite-volume are solved from the eigen-equation of the Hamiltonian matrix, while the final scattering states are calculated by the T-matrix in the infinite-volume.
Finally, we show the black solid and  red dashed lines for the $P^V$ and $P$ defined in Eq.(\ref{eq:PEL}\,, \ref{eq:PE}). 
The $\Delta$ resonance peak is very clean.
This diagram does not show the relationship between the eigenstate of finite-volume and the resonance, but also shows that the $P^V$ agrees very well with $P$ when $L$ goes to infinite.
Furthermore, for the small lattice size, you can find there are only several eigenstates around the peak area, it suggests that these states reflect the properties of the resonance.
Thus, the complements of these eigenstates should reflect the inside structure of resonance.  

Unfortunately, current Lattice techniques  can not measure all eigenstates and eigenvectors for baryon resonances.
At present, for the Baryon spectrum, pure three-quark local operators are mostly used.
Through these operators, we still can have a glance of the inside structure of the excited nucleons.
Overlap with the meson baryon scattering states is suppressed by factor of thousand relative to the nucleon. 
Therefore, the three-quark operator excites the bare state of the Hamiltonian model.
The eigenstates seen on the lattice should have large bare state components.

\subsection{The approach of HEFT}

We brief introduce the approach of HEFT as follows:

\begin{itemize}
\item[1] Fitting experimental data to fix the parameters of the interaction Hamiltonian.\\
\item[2] Using these fitted parameters to generate the Finite-Volume Hamiltonian matrix. And for high pion mass, another parameter for the mass slope is needed.\\
\item[3] Calculate the eigenvalue and eigenvector of this matrix to compare with Lattice data.
\end{itemize}

\section{Results}
\label{sec:results}

In this section, we review the main results of the $N^*(1535)$, $\Lambda^*(1405)$ and $N^*(1440)$ resonances by using HEFT. 

\subsection{For $N^*(1535)$}

\begin{figure*}[tbp]
\begin{center}
\subfigure[]{\includegraphics[width=0.7\columnwidth]{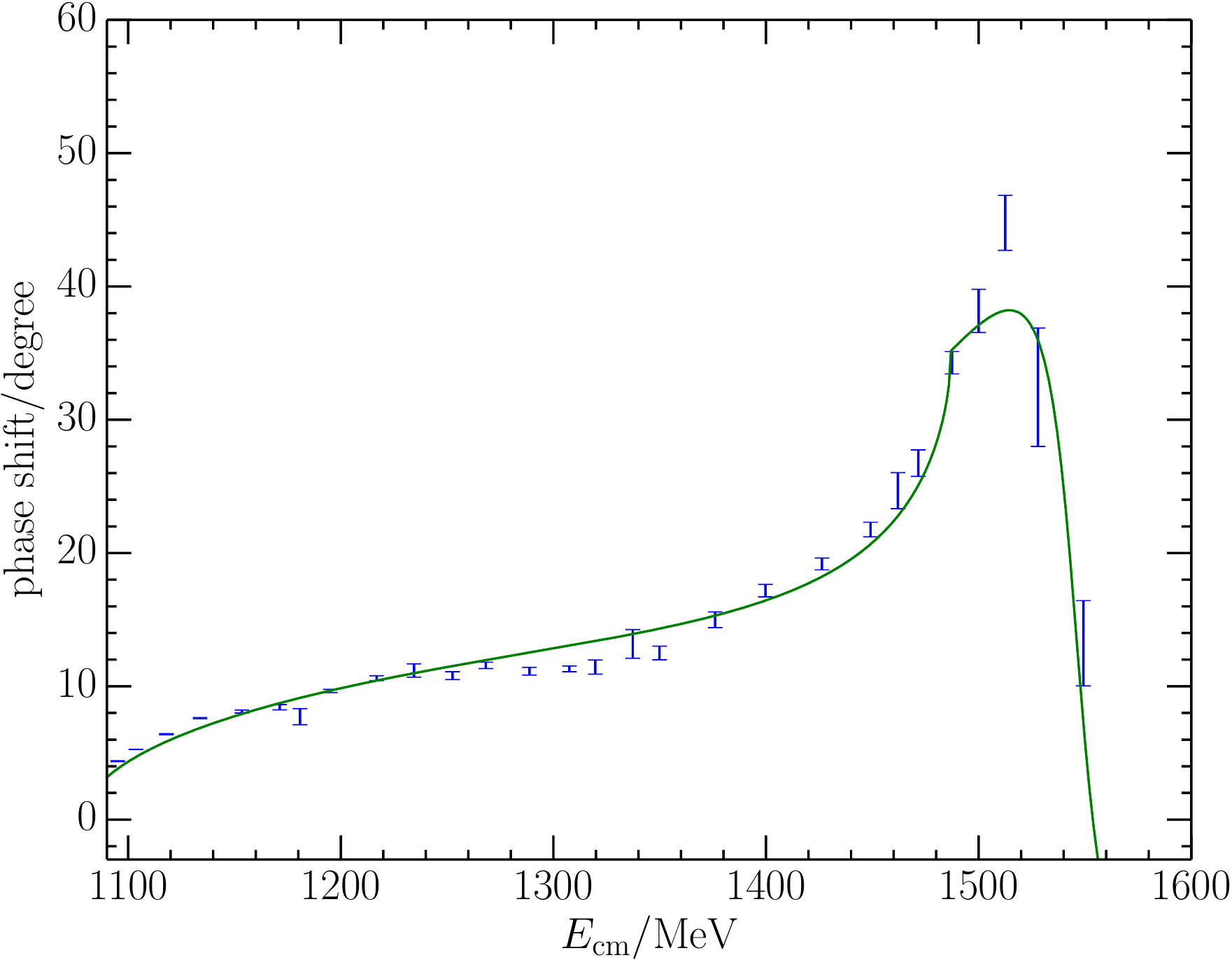}}
\subfigure[]{\includegraphics[width=0.7\columnwidth]{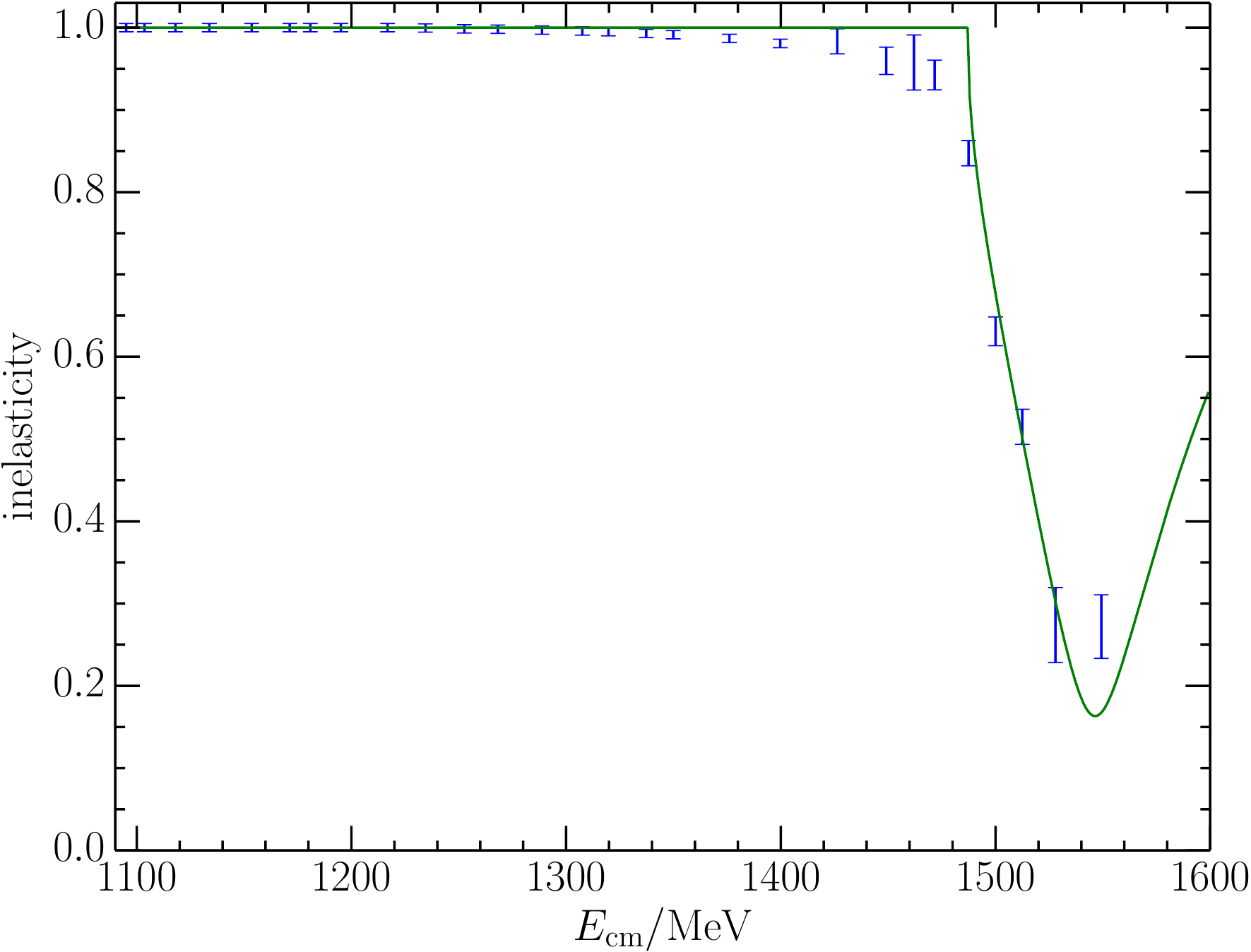}}
\caption{{\bf Colour online:} Experimental data for
the phase shift
  (a) and inelasticity (b) for $\pi N$ scattering with
  $J^P=1/2^-$ are fit by the Hamiltonian model.
  }
\label{fg:fit1535}
\end{center}
\end{figure*}

\begin{figure*}[tbp]
\begin{center}
\subfigure[]{\includegraphics[width=0.7\columnwidth]{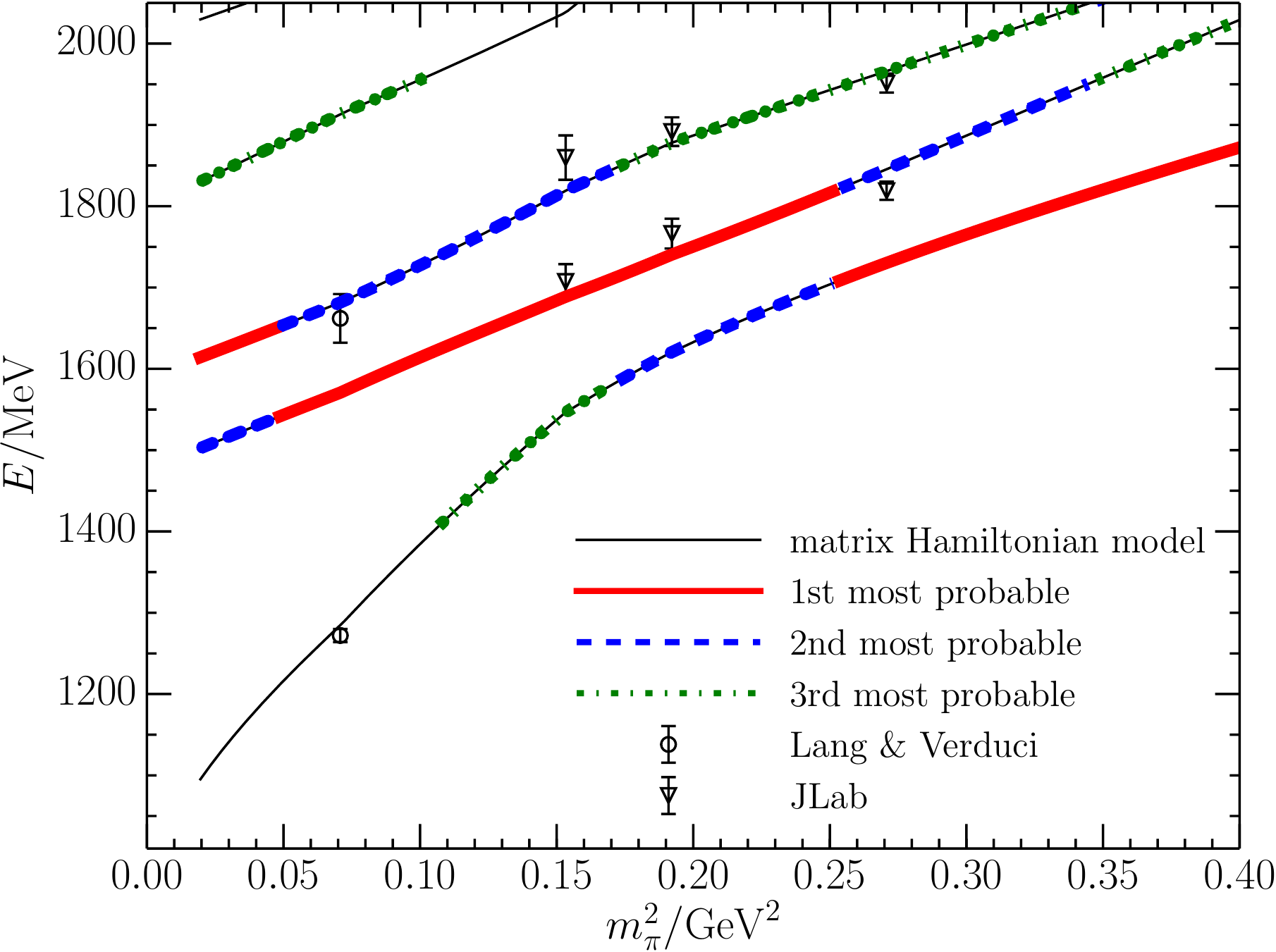}}
\subfigure[]{\includegraphics[width=0.7\columnwidth]{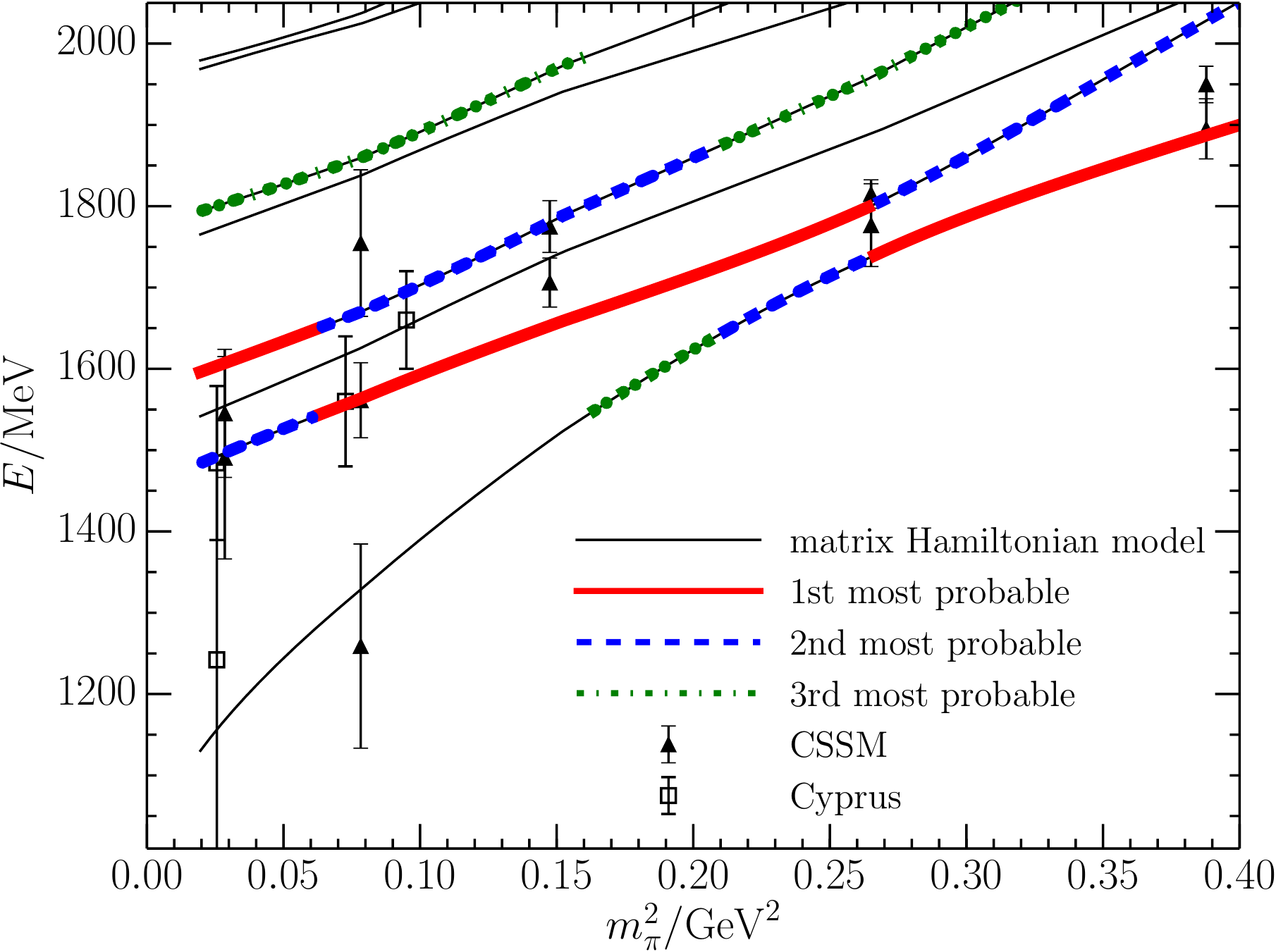}}
\caption{{\bf Colour online:} The pion mass dependence of the $L
  \simeq 1.98$ fm (a) and $L \simeq 2.90$ fm (b) finite-volume
  energy eigenstates.  The different line types and colours 
  indicate the strength of the bare
  basis state in the Hamiltonian-model eigenvector.
  The lattice QCD data are from Jlab~\cite{Edwards:2011jj,Edwards:2012fx}, Lang \& Verduci~\cite{Lang:2012db},
  CSSM~\cite{Kiratidis:2015vpa,Mahbub:2013ala,Mahbub:2012ri} and Cyprus~\cite{Alexandrou:2014mka}.
  }
\label{fg:lat1535}
\end{center} 
\end{figure*}

With the excellent fit of phase shift and inelasticity of $\pi N \to \pi N$ around the $N^*(1535)$ region, 
the parameters $g_{B,\,\pi N}$, $g_{B,\,\eta N}$ and $g_{\pi N,\,\pi N}$ in the Hamiltonian are fixed.
The fits of phase shift and inelasticity are shown in Fig.\ref{fg:fit1535}.
Then the bare mass slope is fixed by fitting various lattice QCD data,and we get the spectra with $L\sim 2$ and $3$ fm, as shown in Fig.\ref{fg:lat1535}.
The values of fitted parameters are listed in Ref.~\cite{Liu:2015ktc}.
Furthermore, it also shows the three eigenstates with largest bare state components as red, blue and green colors.
The lattice QCD data following the color lines is well consistent with the assumption that three-quark operators couple most strongly to this bare state component.
It indicates that our explanation based on HEFT agrees with lattice QCD results.
Typically, the Hamiltonian eigenstates around $1535$ MeV at the physical $\pi$ mass region are dominated by bare-state contributions.
It reflects that the main component of $N^*(1535)$ is the bare state, i.e, three-quark core.
Therefore, the $N^*(1535)$ as the first negative-parity excitation of the nucleon, almost 600 MeV higher in mass is following the expectation based upon the harmonic oscillator model.
We should point out it does not mean $N^*(1535)$ is a pure three-quark state, from our calculation, the fourth eigenstate also includes the components from $\pi N$ and $\eta N$ scattering state. 
But that is no doubt that $N^*(1535)$ is dominated by the three-quark state.

\subsection{For $\Lambda^*(1405)$}

\begin{figure*}[tbp]
\begin{center}
\subfigure[\ $K^-p\to K^-p$]{\includegraphics[width=0.49\columnwidth]{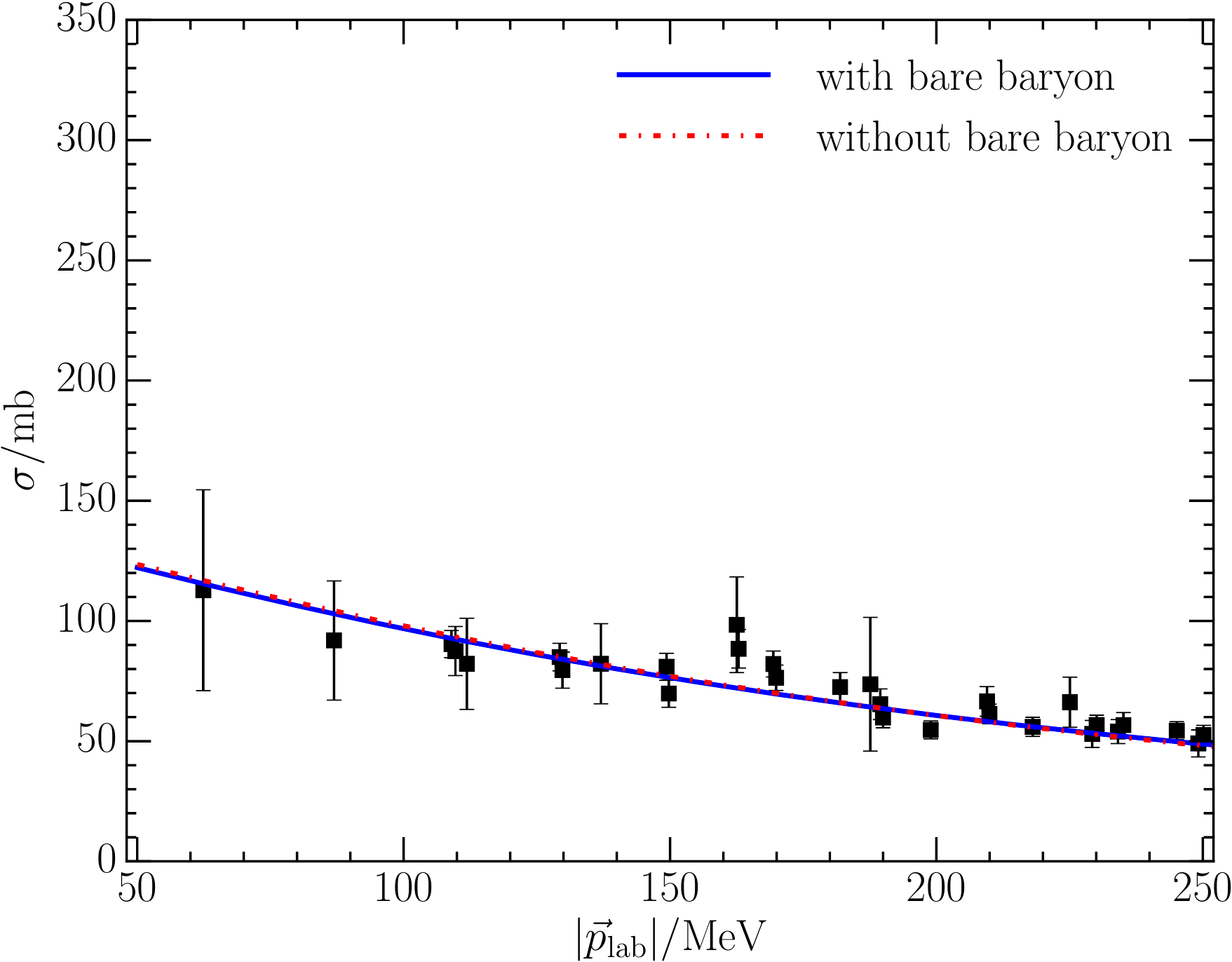}}
\subfigure[\ $K^-p\to\bar K^0n$]{\includegraphics[width=0.49\columnwidth]{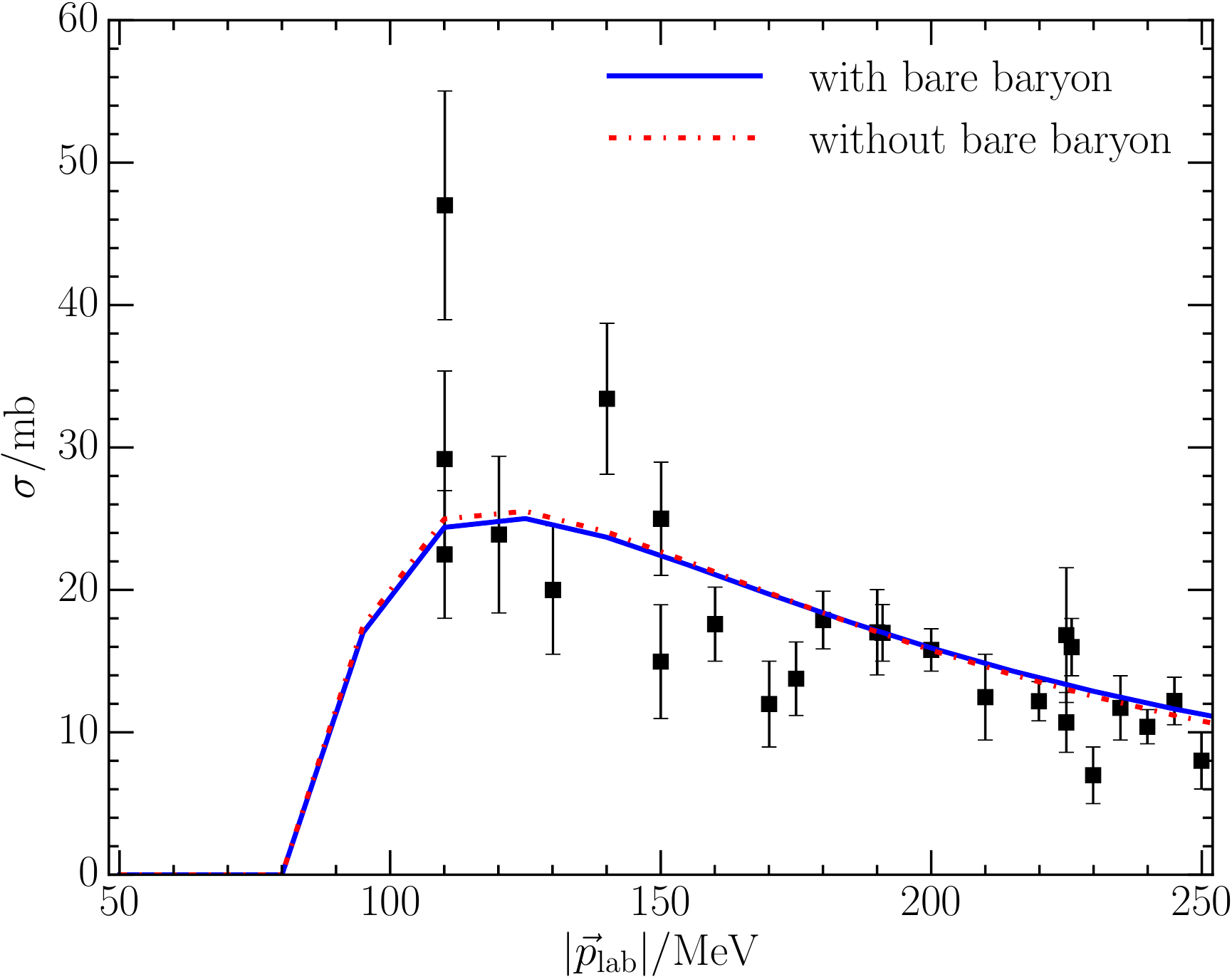}}
\subfigure[\ $K^-p\to \pi^-\Sigma^+$]{\includegraphics[width=0.49\columnwidth]{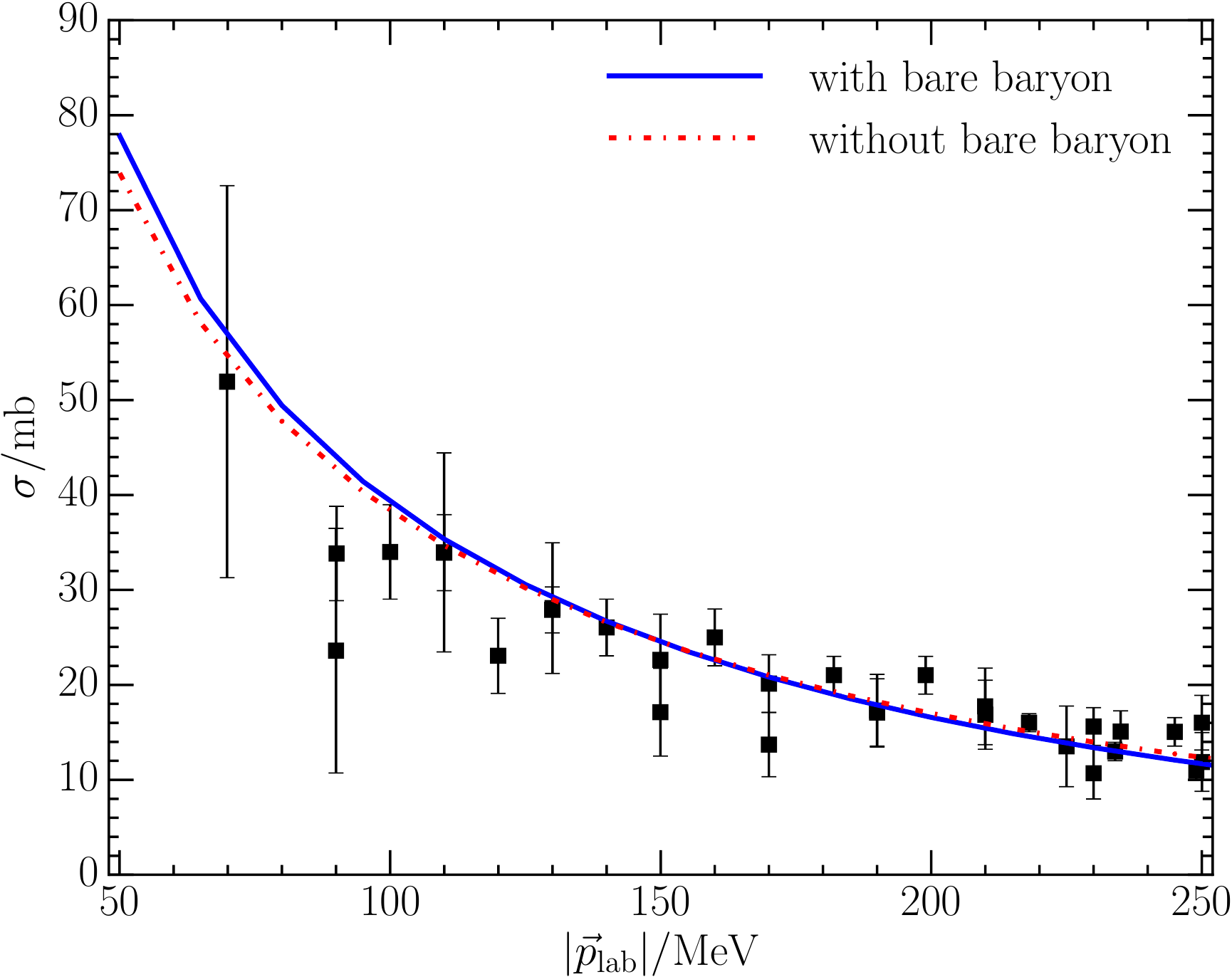}}
\subfigure[\ $K^-p\to \pi^0\Sigma^0$]{\includegraphics[width=0.49\columnwidth]{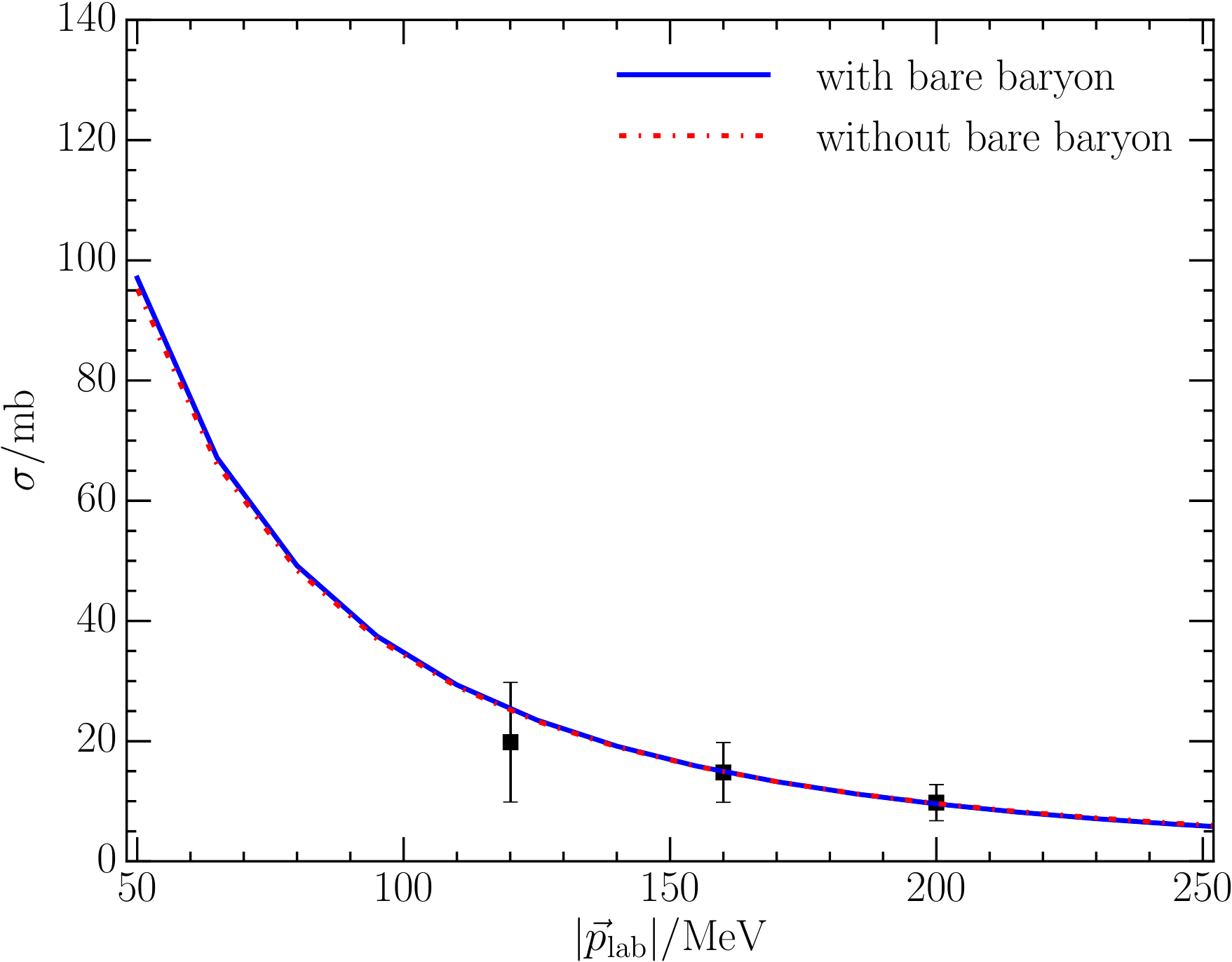}}
\subfigure[\ $K^-p\to \pi^+\Sigma^-$]{\includegraphics[width=0.49\columnwidth]{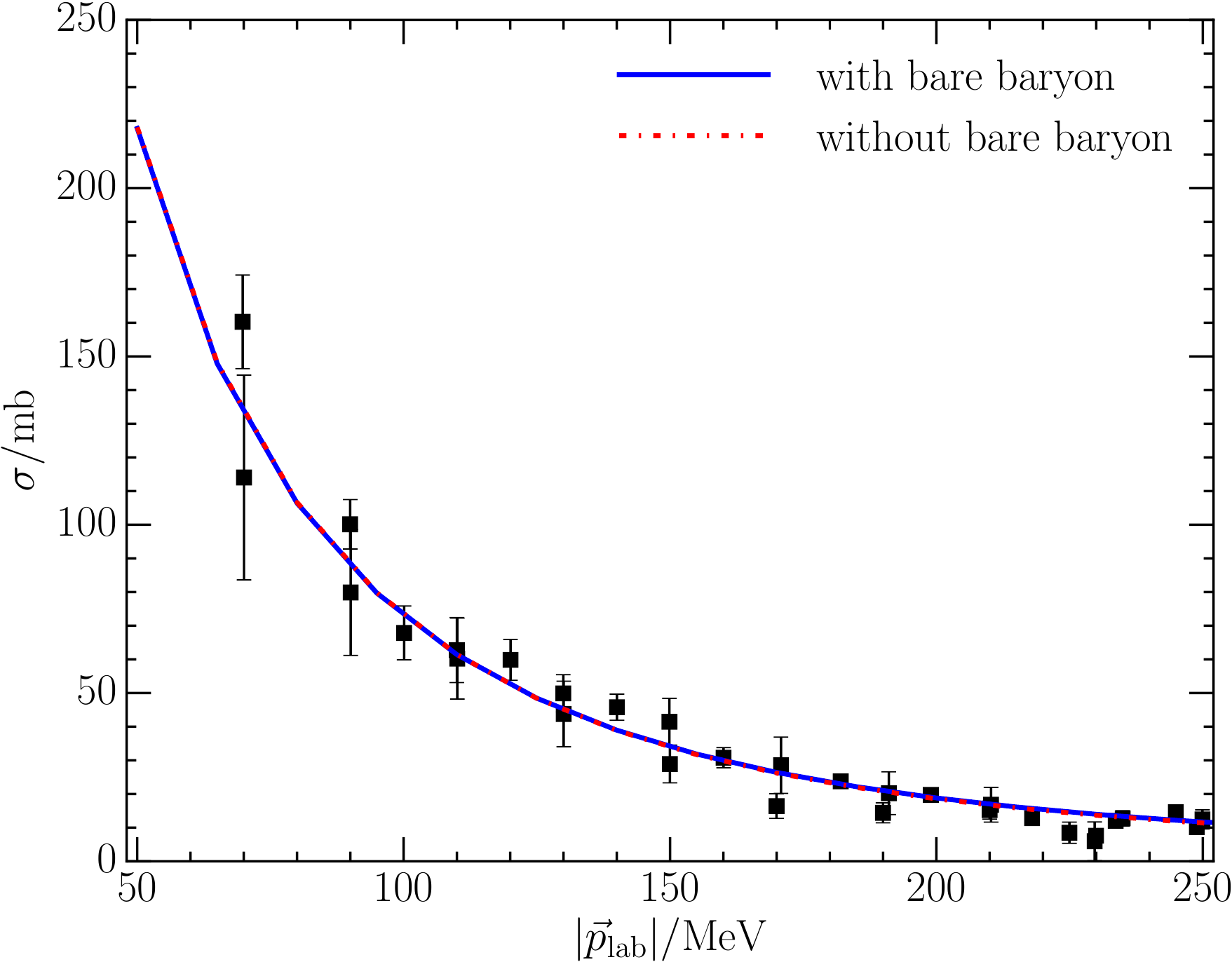}}
\subfigure[\ $K^-p\to \pi^0 \Lambda$]{\includegraphics[width=0.49\columnwidth]{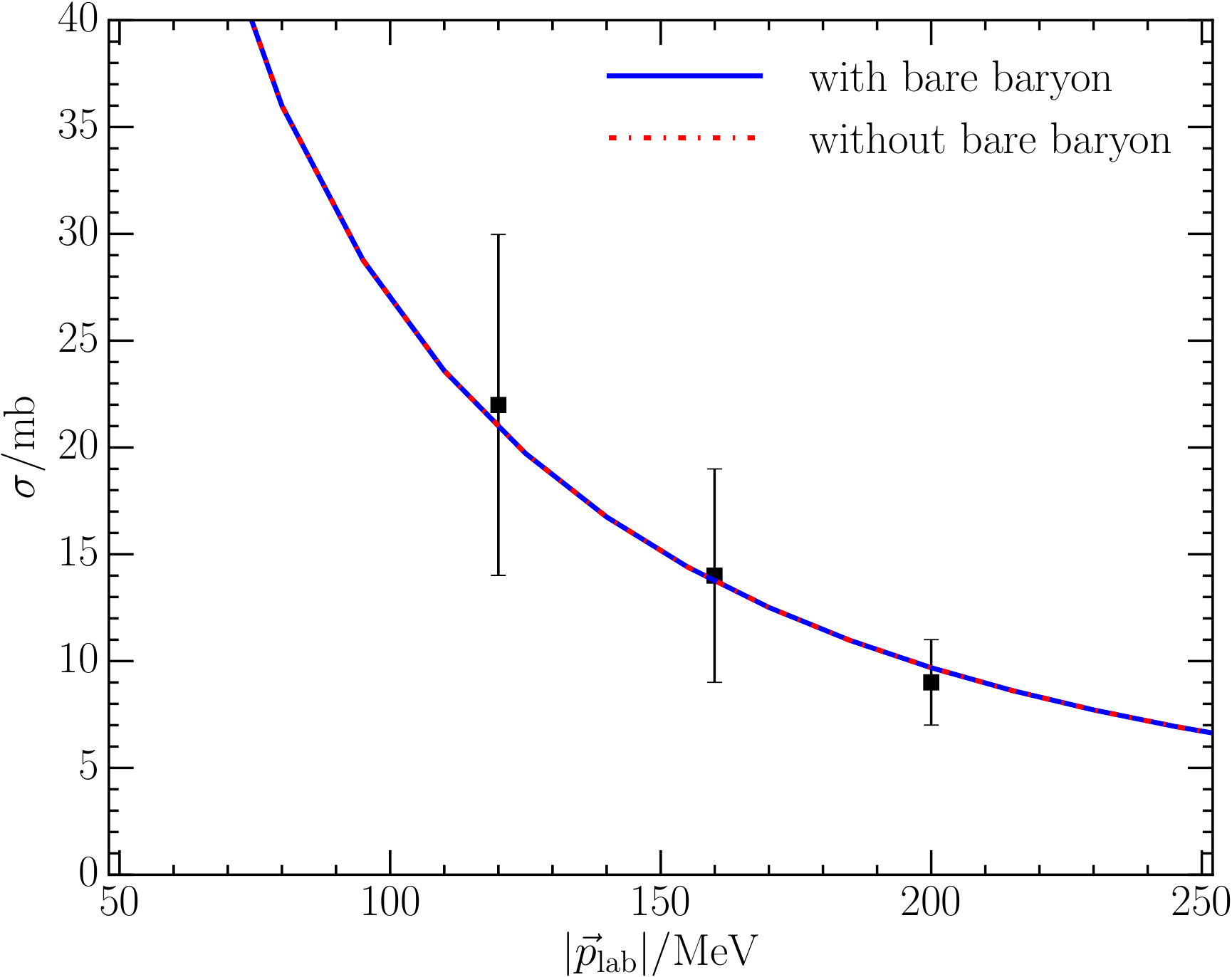}}
\caption{Experimental data and our fits to the cross sections of $K^- p$.  The solid lines are for
  our scenario with a bare-baryon component included in the $I=0$ channel, and the dashed lines
  represent the results without a bare-baryon component.  The experimental data are from
  Refs.~\cite{Sakitt:1965kh,Kim:1965zzd,Csejthey-Barth:1965izu,Mast:1975pv,Bangerter:1980px,Ciborowski:1982et,Evans:1983hz}.}
\label{fg:fit1405}
\end{center} 
\end{figure*}

\begin{figure}[tbp]
\begin{center}
\subfigure[]{\includegraphics[width=0.7\columnwidth]{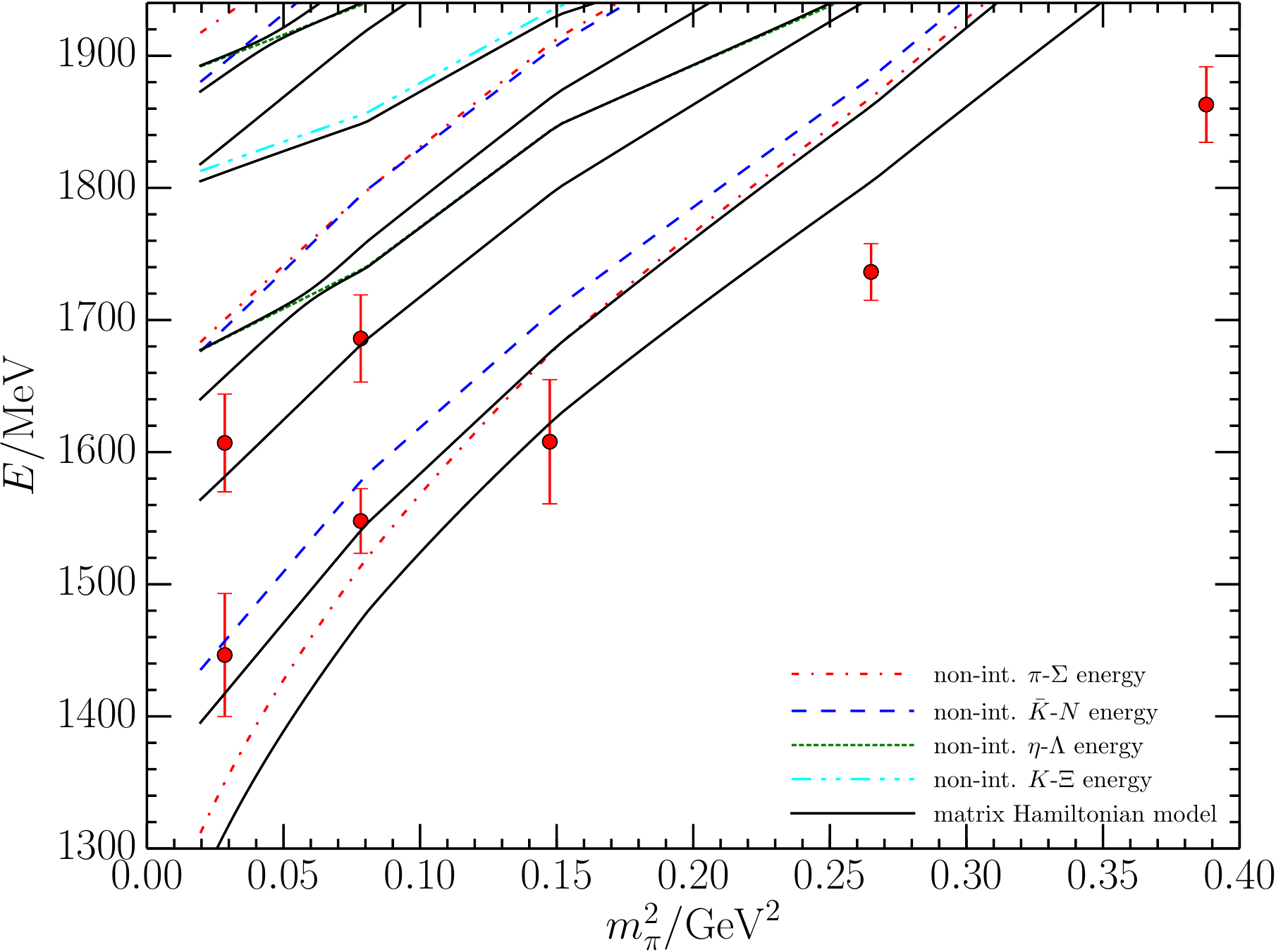}}
\subfigure[]{\includegraphics[width=0.7\columnwidth]{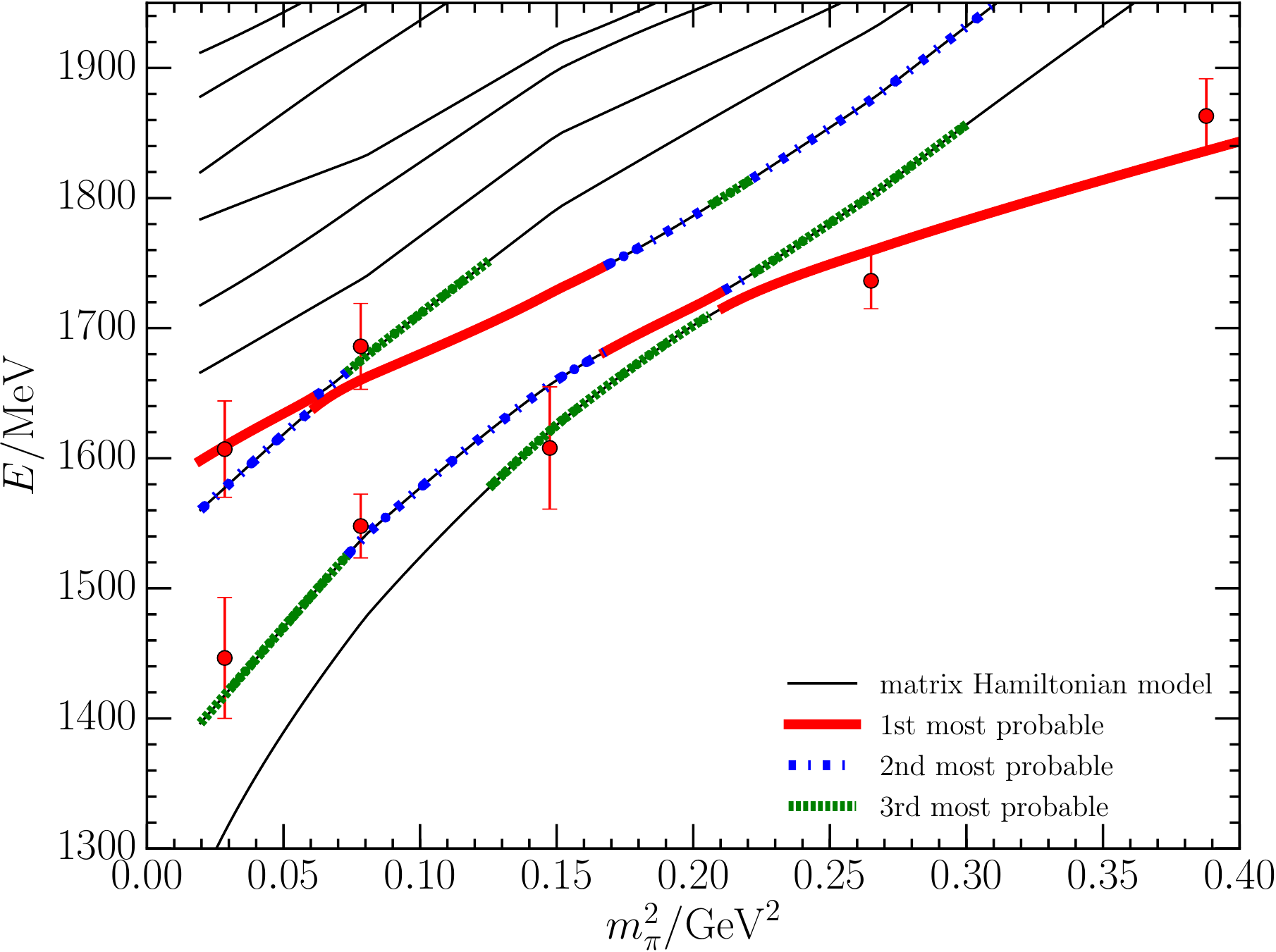}}
\caption{{\bf Colour online:} The pion-mass dependence of the finite-volume energy eigenstates for
  the scenario without(a) and with(b) a bare-baryon basis state.
  The broken lines represent the non-interacting
  meson-baryon energies and the solid lines represent the spectrum derived from the matrix
  Hamiltonian model.  
  The thick-solid (red), dashed (blue) and short-dashed (green)
  lines correspond to the first, second, and third strongest bare-state contributions, and
  therefore the most likely states to be observed with three-quark interpolating fields.
  The lattice QCD results are from the CSSM \cite{Hall:2014uca,Menadue:2011pd}.}
\label{fg:lat1405}
\end{center}
\end{figure}

\begin{figure*}[tbp]
\begin{center}
\subfigure[\ 2nd eigenstate ]{\includegraphics[width=0.7\columnwidth]{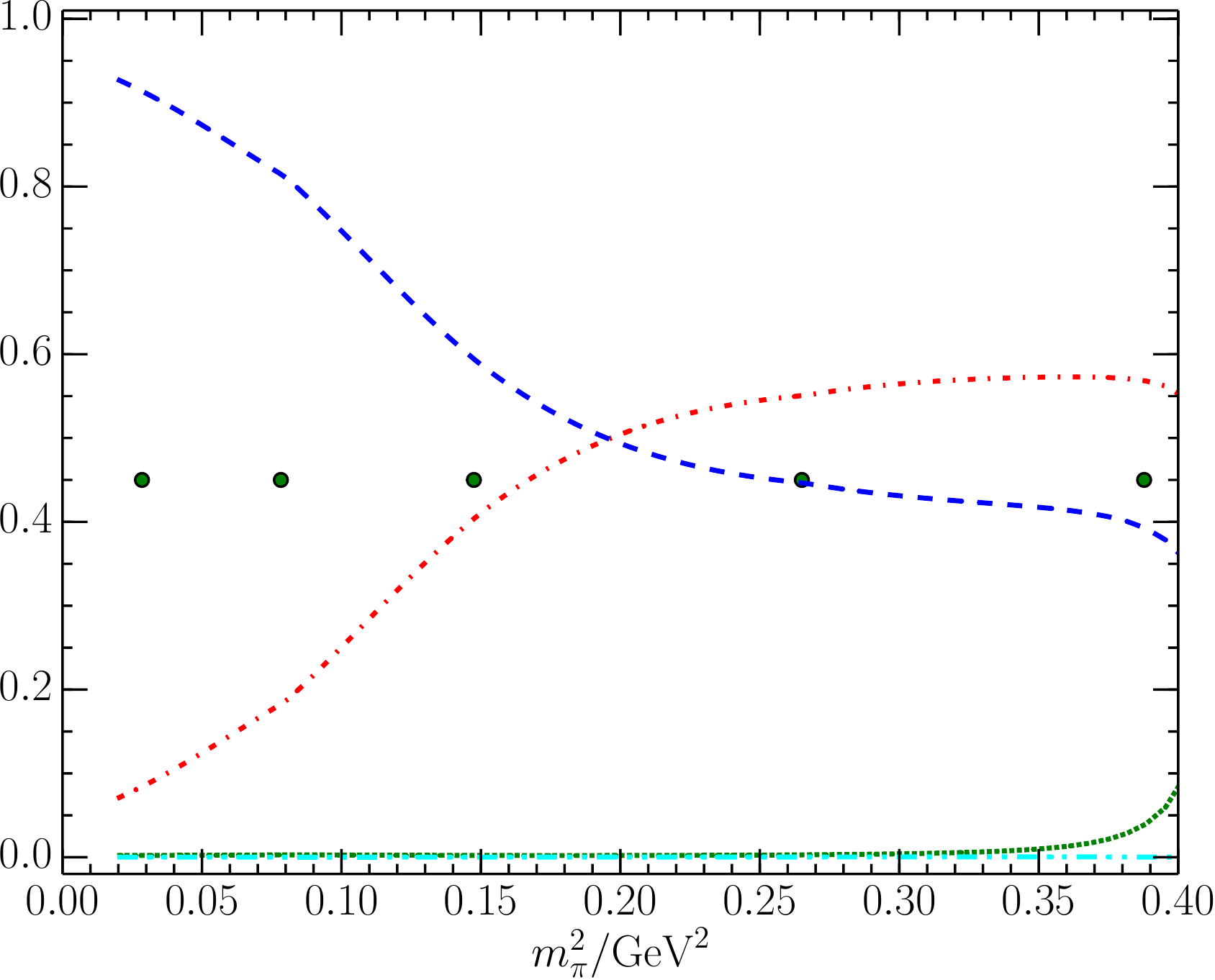}}
\subfigure[\ 2nd eigenstate]{\includegraphics[width=0.7\columnwidth]{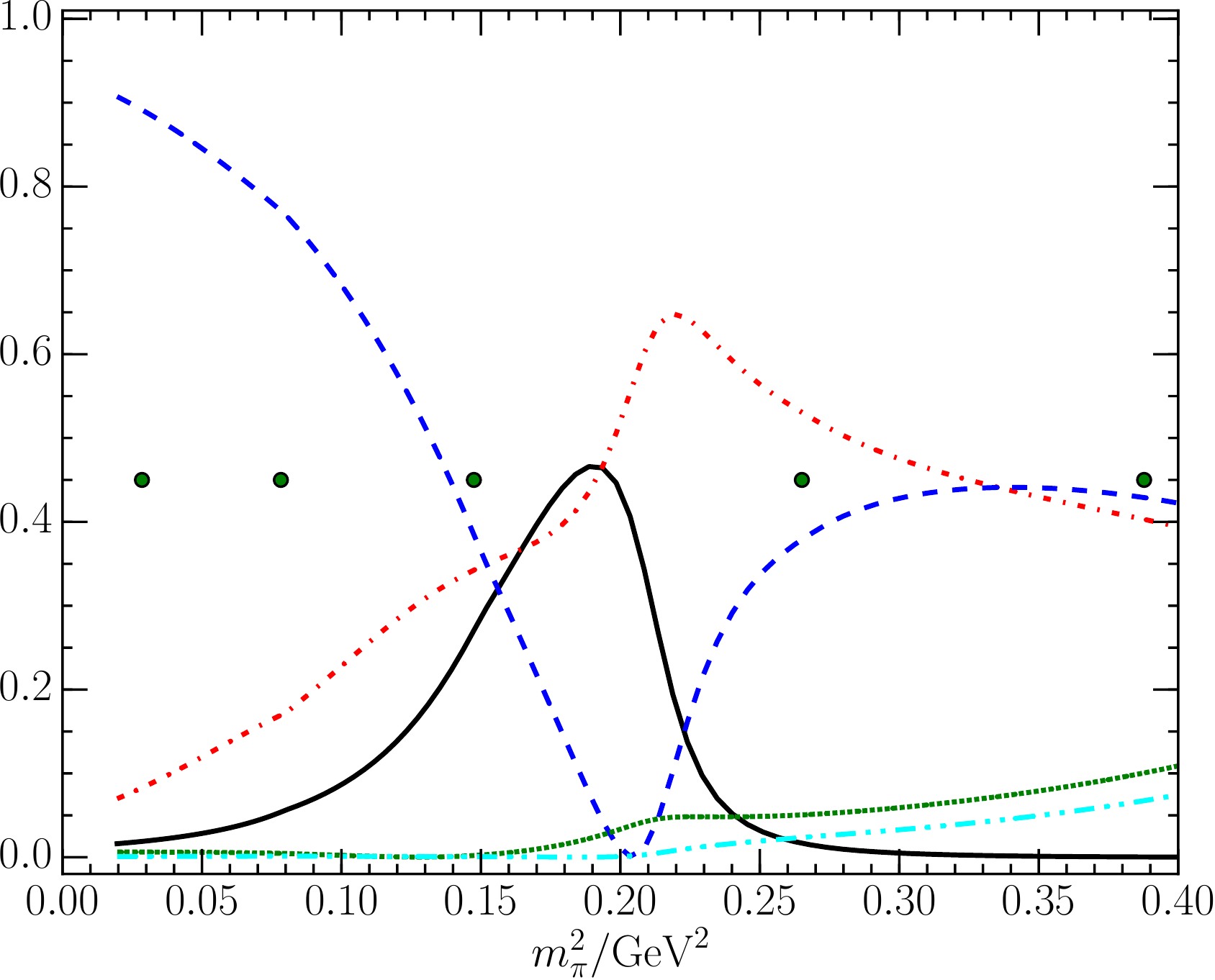}}
\subfigure[\ 4th eigenstate]{\includegraphics[width=0.7\columnwidth]{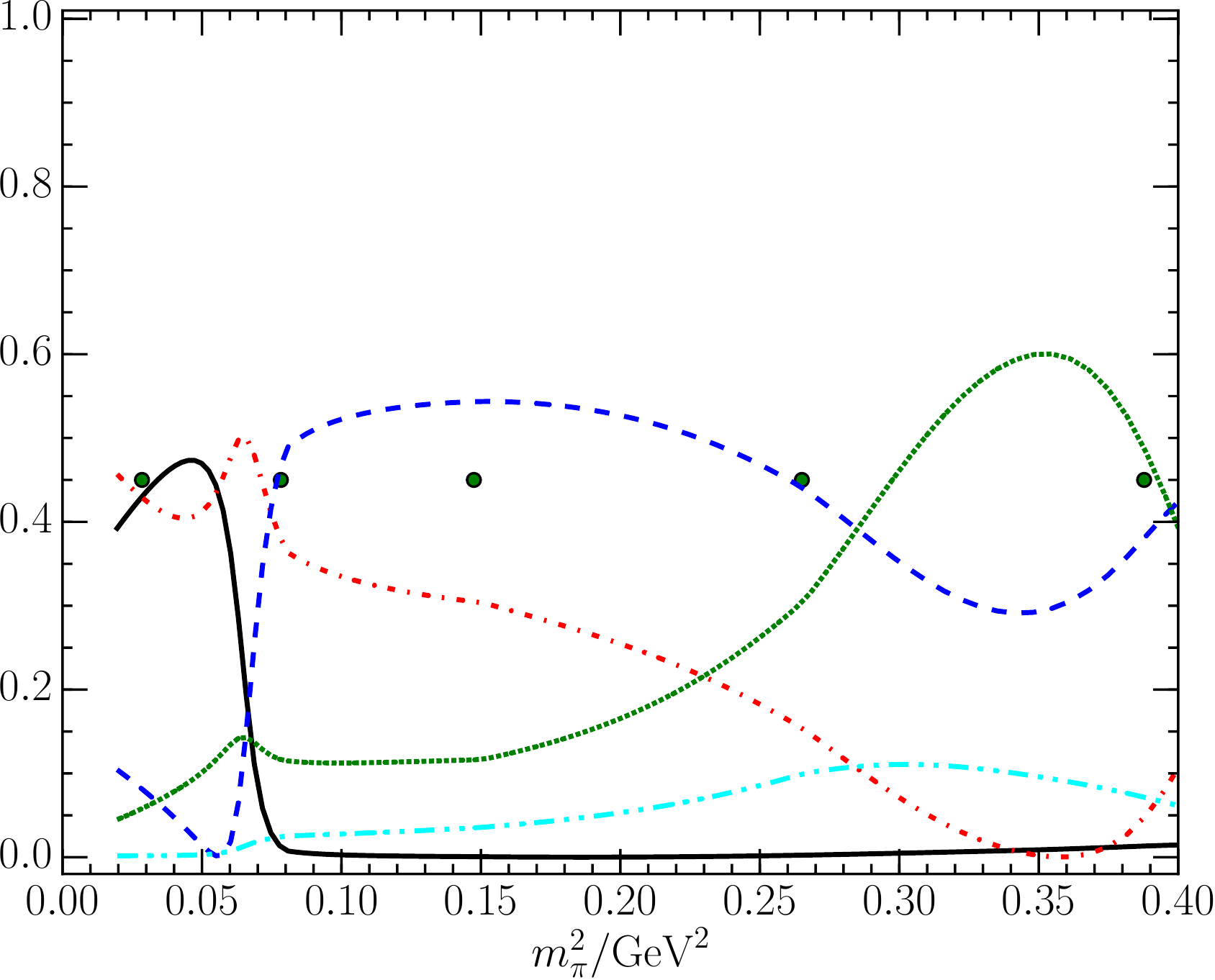}}
\caption{{\bf Colour online:} The pion-mass evolution of the Hamiltonian eigenvector components for
  the eigenstates observed in the scenario. The (a) and (b) are for the second eigenstate without and with bare state scenario. The (c) is for the fourth eigenstate with bare state scenario.  
  The black solid, red dot-dashed, blue dashed, green dotted and cyan dot-dot-dashed lines are for bare state, $|\pi N\rangle$, $|\bar{K} N\rangle$, $|\eta\Lambda\rangle$ and $|K \Xi\rangle$, respectively.
  The (green) dots plotted
  horizontally at $y = 0.45$ indicate the positions of the five quark masses considered by the CSSM
  on a lattice volume with $L\simeq 2.90$ fm.}
\label{fg:dis1405}
\end{center}
\end{figure*}

For the $\Lambda^*(1405)$ case, the cross sections of various channels are fitted as shown in Fig.\ref{fg:fit1405}.
Here, two different models with and without bare state are developed.
The fitted parameters values are shown in Ref.~\cite{Liu:2016wxq}.
The corresponding two spectra are shown in Fig.\ref{fg:lat1405}.
Furthermore, the various components of second and fourth eigenstates of the finite-volume Hamiltonian matrix are shown in Fig.\ref{fg:dis1405}.
It is clear that the two lattice QCD points at the largest $\pi$ mass can be only explained by the model with bare state.
However, it is interesting to notice that at the physical $\pi$ mass region, even in the with bare state model, the second eigenstate around $1405$ MeV just takes around $5\%$ bare state components.
The components of the second eigenstate in the two models are both dominated by $\bar{K} N$ components as shown in Fig.\ref{fg:dis1405}.
It is confirmed that the main components of $\Lambda^*(1405)$ is the $\bar{K} N$.
This result indicates that $\Lambda^*(1405)$ can not be recognize as a three-quark state, while it is better to described as a bound state of $\bar{K} N$ near the physical pion mass. 
It is worthwhile to point out that the fourth eigenstate, having the largest bare state component is above 1600 MeV.
It is also consistent the prediction of the harmonic oscillator model for the first excitation of $\Lambda$ resonance. 

\subsection{For $N^*(1440)$}

\begin{figure}[tbp]
\begin{center}
\includegraphics[width=1.\columnwidth]{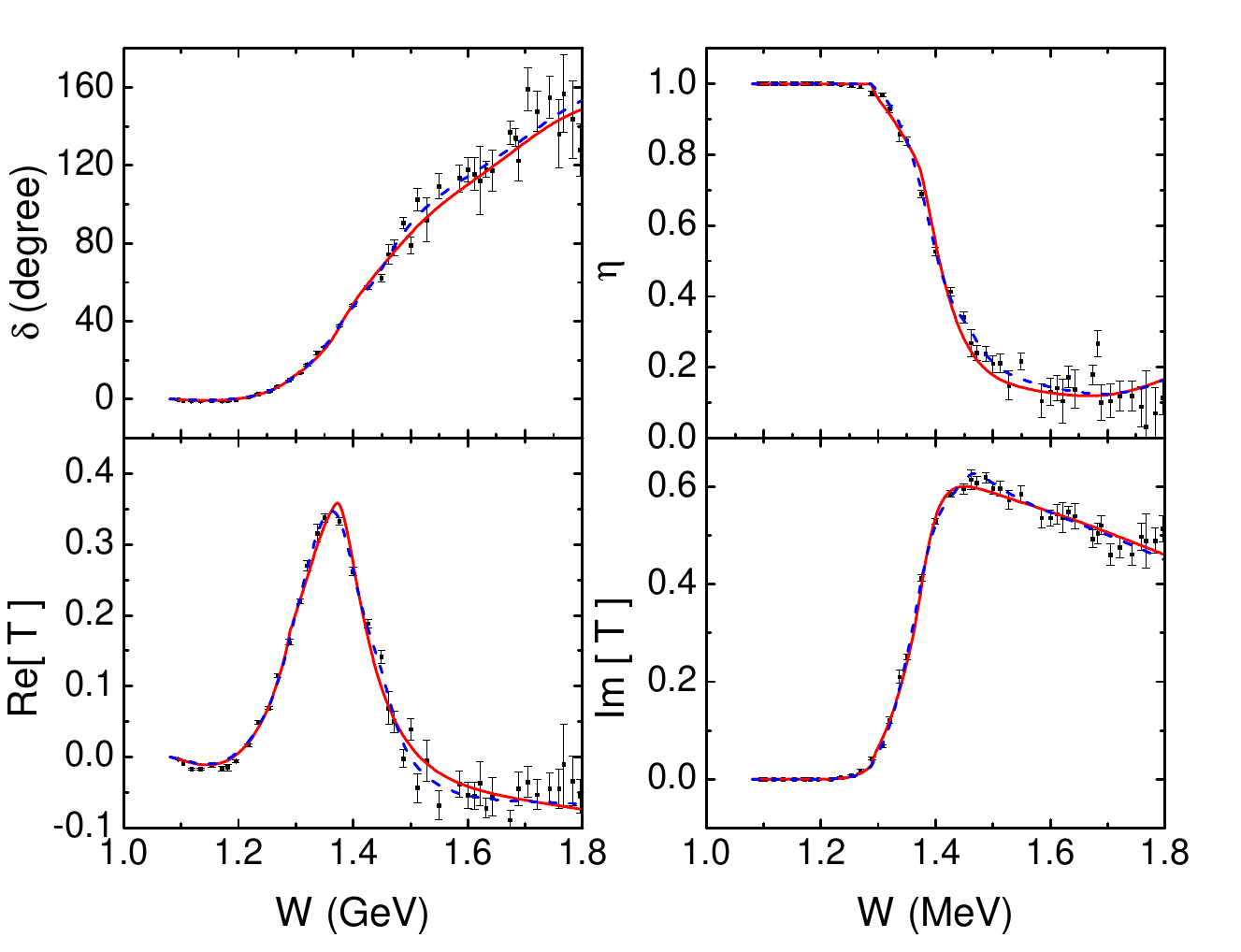}
\caption{
The fitted phase shift $\delta$, inelasticity $\eta$ and $T$-matrix
for the $\pi N \to \pi N$ reaction. Red-solid and blue-dashed lines
are calculated from scenarios I and II, respectively.
}\label{fg:fit1440}
\end{center}
\end{figure}

There is a long puzzle for the nature of the Roper, $N^*(1440)$.
By a similar method as that applied for the $N^*(1535)$ and $\Lambda^*(1405)$ resonances, the $N^*(1440)$ is studied with the experimental data and recent lattice QCD results.
The fitted phase shift, inelasticity and T-matrix for $\pi N \to \pi N$ are shown in Fig.\ref{fg:fit1440}.
We find two rather different models which appear equally acceptable in terms of the quality of the fit to existing data, as red-solid and blue-dashed lines in  Fig.\ref{fg:fit1440}.
The values of parameters are listed in Ref.~\cite{Wu:2017qve}.
The main differences between these two models are the coupling of $\pi N \to \pi N$ in scenario I is much larger than another, while the coupling of the bare state to $\pi N$ and $\sigma N$ in scenario I is much smaller.
These differences leads two pictures of the Roper, it is generated by the strong rescattering in the baryon-meson channels or dominated by the bare state.

\begin{figure}[tbp]
\begin{center}
\subfigure[]{\includegraphics[width=0.7\columnwidth]{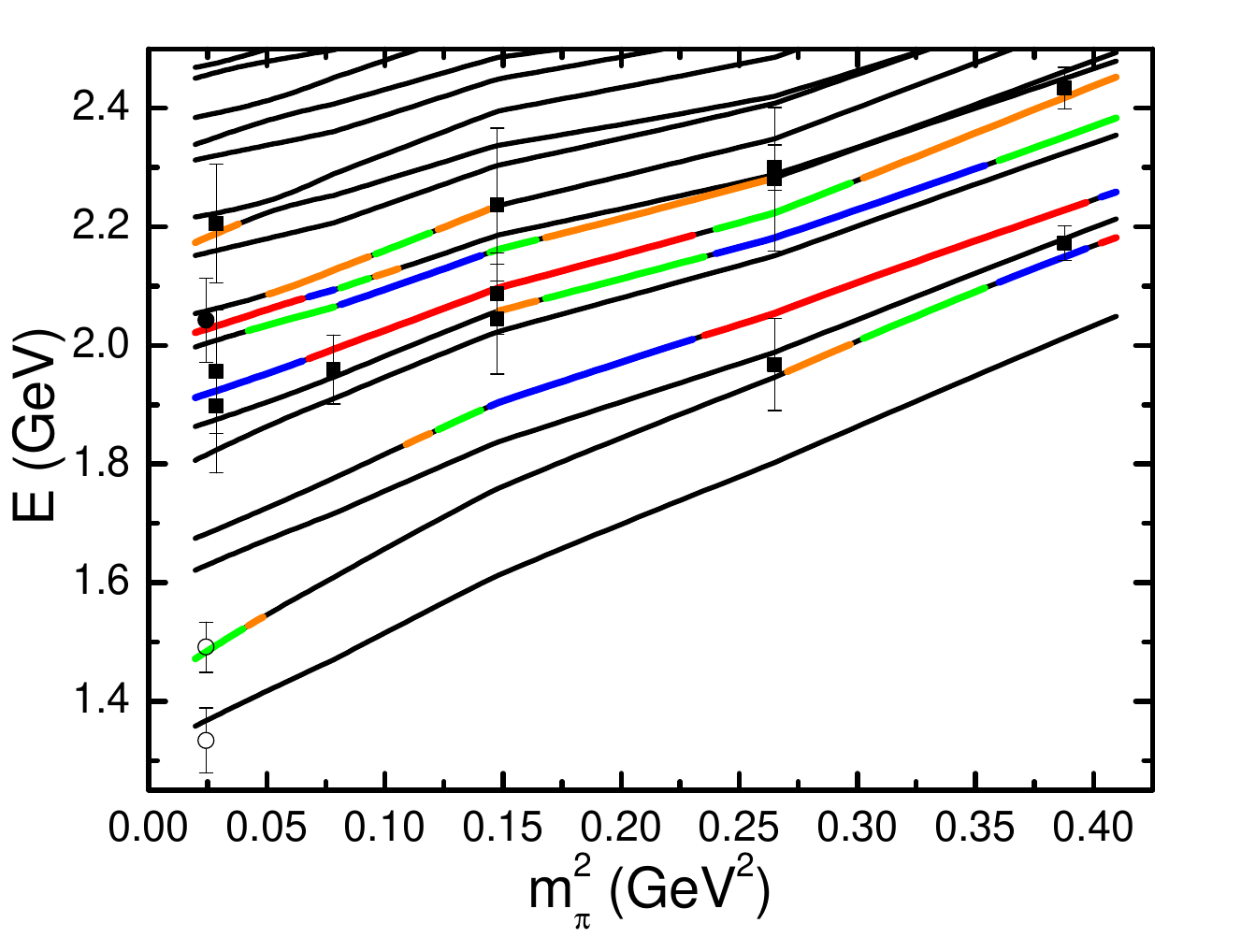}}
\subfigure[]{\includegraphics[width=0.7\columnwidth]{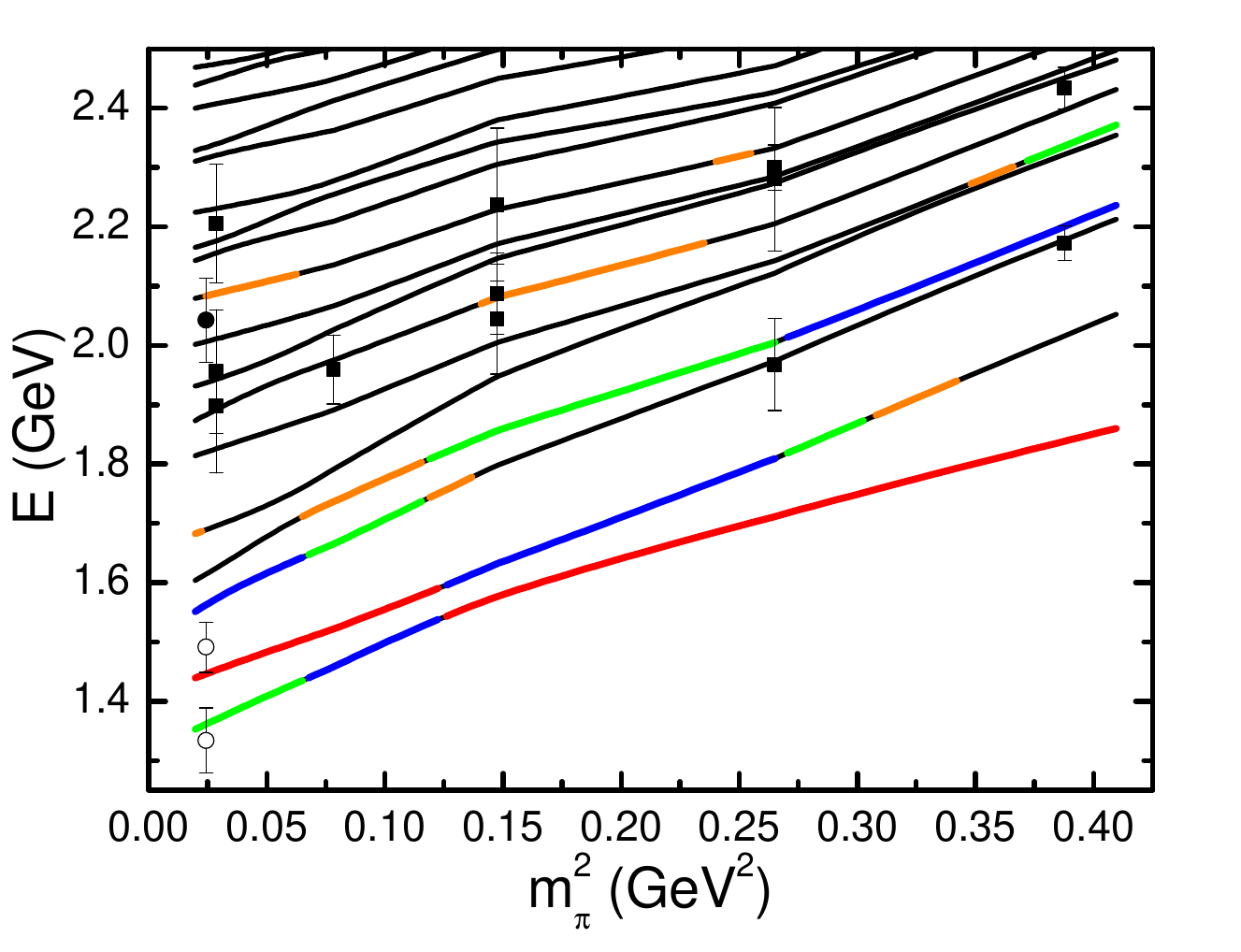}}
\caption{
The finite volume spectrum of Scenario I(a) with a bare mass of 2.0 GeV and Scenario II(b) having a bare mass of 1.7 GeV.
The CSSM results \cite{Liu:2016uzk} are indicated by square symbols
and circles denote the more recent results from Lang {\it et al.}
\cite{Lang:2016hnn}.  Solid symbols indicate states dominated by local
three-quark operators and open symbols indicate states dominated by
non-local momentum-projected five-quark operators.  The colours red,
blue, green and orange are used to indicate the relative contributions
of the bare baryon basis state in the eigenstate, with red being the
largest contribution.
}\label{fg:lat1440}
\end{center}
  \end{figure}

Through HEFT approach, two spectra in the finite volume are shown in Fig.\ref{fg:lat1440}.
The predictions of scenario I are consistent with lattice QCD because all of the lattice states dominated by local three-quark interpolating fields can be associated with a colored line.  
For example, all of the Hamiltonian states having the largest bare basis-state component, indicated in red in Fig.~\ref{fg:lat1440}, have a nearby lattice QCD result.  
On the other hand, scenario II displays little correspondence to the lattice QCD results. 
Scenario II predicts a low-lying state with a large bare basis-state component of approximately 50\%, approaching that for the ground state.  
Such a state would be easy to excite in lattice QCD with local three-quark operators.  
However this state is not seen in the simulations through the local three-quark operators. 
Indeed,  Lang {\it et al.} \cite{Lang:2016hnn} only see this state when they include a non-local $\pi N$ interpolating field.
Obviously, scenario II fails to explain lattice data.
Furthermore, the lowest two states observed by the Lang {\it et al.} group are quite consistent with scenario I.
Near the physical mass the first eigenstate is dominated by $\sigma N$ basis states, while the main components of the second state is $\pi N$ basis states.
The Lang group find only the first and second state when they include a non-local $\sigma N$ and $\pi N$ interpolating field, respectively.
From this study of the Roper, the mystery of the low-lying Roper resonance may be nearing resolution.
Evidence indicates the observed nucleon resonance at 1440 MeV is best described as the result of strong rescattering between coupled meson-baryon channels.
And the excited state from three-quark core is around 1.9 GeV, which is consistent with the harmonic oscillator model.

\subsection{The picture of the lowest $N^*$ and $\Lambda^*$}

Through HEFT to study of $N^*(1535)$, $\Lambda^*(1405)$ and $N^*(1440)$, the nature of these resonance states is found as follows:
\begin{itemize}
\item[1] For $N^*(1535)$, it is mainly dominated by a  three-quark core.\\

\item[2] For $\Lambda^*(1405)$, it is predominantly a molecular $\bar{K} N$ bound
state. 
The excited state of the quark model is around 1.6 GeV.\\

\item[3] For $N^*(1440)$, it is best described as the result of strong rescattering between coupled meson-baryon channels.
The first radial excited nucleon of the quark model is around $1.9$ GeV.
\end{itemize}

\begin{figure}[tpb]
\hspace{0.25cm}\includegraphics[width=1.0\columnwidth]{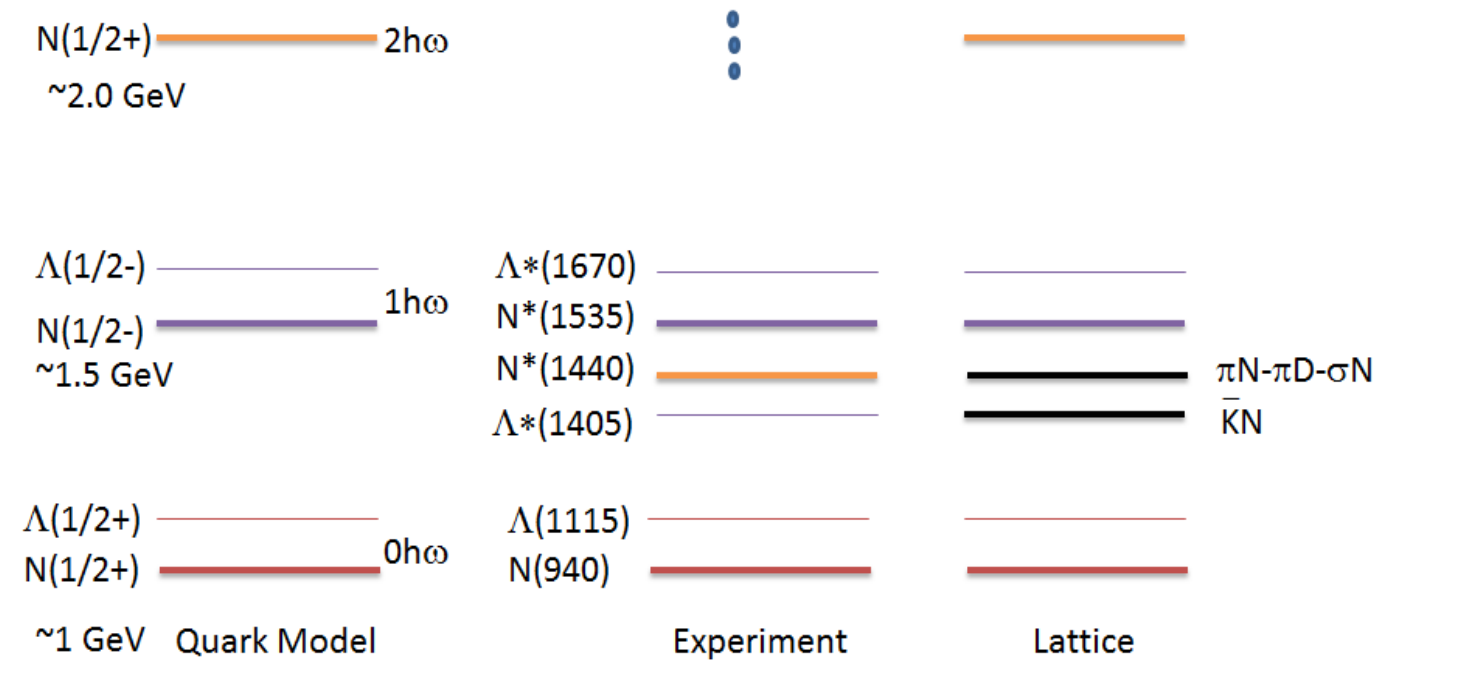}
\caption{
The spectra of $N$ and $\Lambda$ of the quark model, experimental data and the analysis of this paper.
}\label{fg:newp}
  \end{figure}

There conclusions lead to the new picture of low lying $N$ and $\Lambda$ resonances as shown in Fig.\ref{fg:newp}.
From the quark model, it predicts three levels of hadron mass.
However, from the experimental data,  $\Lambda^*(1405)$ and $N^*(1440)$ do not obey this mass order.
Through the analyses of Lattice QCD results based on HEFT, the nature of  $\Lambda^*(1405)$ and $N^*(1440)$ are predominantly meson-baryon states,  rather than three-quark states.
In other words, these two state are beyond the quark model.
On the other hand, the states predicted in the quark model all have evidence from the analyses of lattice QCD results.

\section{Conclusion}
\label{sec:con}

In this report, HEFT approach is introduced.
HEFT connects the experimental data, Lattice QCD results and the properties of resonances.
From the Hamiltonian matrix in the finite-volume, it does not only get the eigenvalues which corresponding to the spectrum of Lattice QCD simulation, but also provides the eigenvectors of these eigenstates.
The eigenvectors will be useful to distinguish the different models and reflect the insight of the resonances.
The $N^*(1535)$, $\Lambda^*(1405)$ and $N^*(1440)$  are studied based on HEFT combined with experimental and lattice QCD data.
The lattice QCD data show that the $\Lambda^*(1405)$ and $N^*(1440)$ are dominated by the meson-baryon interaction, while the $N^*(1535)$ is mainly composed of three-quarks.
Furthermore, from lattice QCD data, there is evidence to show that the first excited state of the $\Lambda$ is around $1.6$ GeV, and the first radial excited state of nucleon is around $1.9$ GeV.
All of above conclusions show that the states found in the lattice QCD results are consistent with the expectations of the harmonic oscillator model, except $\Lambda^*(1405)$ and $N^*(1440)$.  

The definitive analysis for the picture of hadron needs new data to further resolve the nature of these resonance states.
It requires new experimental data and lattice QCD simulations, especially, also complete nucleon spectrum on several lattice volumes.




\end{document}